\documentclass[conference]{IEEEtran}
\usepackage{cite}
\usepackage{xurl}

\usepackage{hyperref}
\hypersetup{colorlinks=true, urlcolor=blue, hypertexnames=false} 

\usepackage{tikz}
\usetikzlibrary{positioning,arrows.meta,shapes.geometric,fit,calc,patterns}
\usepackage{amsmath}
\usepackage{graphicx}
\usepackage{booktabs}
\usepackage{array}
\usepackage{colortbl}
\usepackage{multirow}
\usepackage{pifont}
\newcommand{\cmark}{\textcolor{green!60!black}{\ding{51}}}
\newcommand{\xmark}{\textcolor{red!70!black}{\ding{55}}}
\usepackage[most]{tcolorbox}
\usepackage{float}
\usepackage{placeins}
\usepackage{dblfloatfix}
\usepackage{enumitem}
\graphicspath{{figures/}}

\newcounter{subfigure}
\renewcommand{\thesubfigure}{\alph{subfigure}}

\makeatletter
\renewcommand{\p@subfigure}{\thefigure}
\makeatother
\newcommand{\duetbeginsubfigures}{\stepcounter{figure}\setcounter{subfigure}{0}}
\newcommand{\duetfinishsubfigures}{\addtocounter{figure}{-1}}
\newcommand{\duetsubcaption}[1]{%
  \refstepcounter{subfigure}%
  \par\vspace{2pt}\footnotesize\textbf{(\thesubfigure)}~#1\par\normalsize%
}

\newtcolorbox{designinsight}{
    enhanced, colback=blue!4, colframe=blue!55!black,
    boxrule=0pt, leftrule=3pt, arc=2pt,
    left=8pt, right=8pt, top=6pt, bottom=6pt
}

\newtcolorbox{goalsbox}{
    enhanced, colback=orange!5, colframe=orange!60!black,
    boxrule=0pt, leftrule=3pt, arc=2pt,
    left=8pt, right=8pt, top=6pt, bottom=6pt,
    fonttitle=\bfseries
}

\newtcolorbox{takeaways}{
    enhanced, colback=green!5, colframe=green!50!black,
    boxrule=0pt, leftrule=3pt, arc=2pt,
    left=8pt, right=8pt, top=6pt, bottom=6pt,
    fontupper=\small
}

\newtcolorbox{schemabox}[1]{
    enhanced, breakable=false,
    colback=white, colframe=orange!55!black,
    colbacktitle=orange!18, coltitle=orange!55!black,
    fonttitle=\bfseries\small, title={#1},
    boxrule=0.5pt, arc=2pt, titlerule=0pt,
    left=4pt, right=4pt, top=3pt, bottom=3pt
}
\newtcolorbox{casebox}[1]{
    enhanced, breakable=false,
    colback=white, colframe=orange!55!black,
    colbacktitle=orange!18, coltitle=orange!55!black,
    fonttitle=\bfseries\footnotesize, title={#1},
    boxrule=0.5pt, arc=2pt, titlerule=0pt,
    left=3pt, right=3pt, top=2pt, bottom=2pt
}
\newcommand{\hlY}[1]{{\setlength{\fboxsep}{1.2pt}\colorbox{yellow!28}{\strut #1}}}
\newcommand{\hlR}[1]{{\setlength{\fboxsep}{1.2pt}\colorbox{red!12}{\strut #1}}}
\newcommand{\hlG}[1]{{\setlength{\fboxsep}{1.2pt}\colorbox{green!18}{\strut #1}}}

\begin{document}

\title{One Pipeline Does Not Fit All: TAILOR, a Type- and State-Aware Framework for CVE Reproduction}

\author{
\begin{tabular}{@{}c@{\hspace{22pt}}c@{\hspace{22pt}}c@{}}
Ji He & Huang Zhang & Lijie Zheng \\
\normalsize Xidian University & \normalsize Xidian University & \normalsize Xidian University \\
\normalsize jihe@xidian.edu.cn & \normalsize 25031212221@stu.xidian.edu.cn & \normalsize lijzheng@stu.xidian.edu.cn
\end{tabular}
\\[1.2em]
\begin{tabular}{@{}c@{\hspace{48pt}}c@{}}
Lele Zheng & Yulong Shen \\
\normalsize Xidian University & \normalsize Xidian University \\
\normalsize zhenglele@xidian.edu.cn & \normalsize ylshen@mail.xidian.edu.cn
\end{tabular}
}
\hypersetup{
  pdftitle={One Pipeline Does Not Fit All: TAILOR, a Type- and State-Aware Framework for CVE Reproduction},
  pdfauthor={Ji He, Huang Zhang, Lijie Zheng, Lele Zheng, Yulong Shen},
  pdfsubject={Type- and state-aware automated CVE reproduction}
}

\maketitle

\begin{abstract}
Growing vulnerability disclosure and widespread software reuse increase security teams' need for reproducible evidence to diagnose vulnerabilities, validate patches, and build regression tests. Producing such evidence at scale requires automated end-to-end CVE reproduction. Existing methods typically process different CVEs through a uniform pipeline, but differences in runtime form, trigger interfaces, and prerequisite state impose different execution requirements on individual stages, making fixed workflows difficult to adapt to diverse reproduction needs. To address this problem, we present TAILOR, a type- and state-aware multi-agent framework specialized for complex vulnerability reproduction. TAILOR converts static vulnerability information into auditable reproduction evidence and packages reconstructed environments and trigger evidence into reproduction artifacts. Its first-level type-aware mechanism adaptively matches each vulnerability to an execution path. Within the Web path, its second-level state-aware mechanism constructs the required prerequisite state before exploitation, decouples prerequisite-state construction from core vulnerability triggering, and shares execution constraints across exploitation and verification. We construct a dataset of 200 CVEs with an emphasis on cases with complex execution requirements. TAILOR successfully reproduces 59.24\% of Web vulnerabilities and 44.19\% of traditional vulnerabilities. Further ablation experiments show that the two control levels respectively mitigate execution-path mismatch and missing Web prerequisite state. Overall, TAILOR broadens the coverage of automated CVE reproduction and provides auditable evidence for vulnerability diagnosis and defense.
\end{abstract}

\section{Introduction}

As modern software systems and their dependency ecosystems continue to expand, the number of publicly disclosed vulnerabilities is growing rapidly~\cite{sabottke2015disclosure}. The U.S. National Institute of Standards and Technology (NIST) reports that CVE submissions increased by 263\% from 2020 to 2025, and existing vulnerability databases can no longer keep pace with the continuing growth of incoming reports~\cite{nist_nvd_growth_2026}. Modern applications typically consist of a large number of third-party libraries and services. A dependency vulnerability that is not confirmed and handled in time can further expand the attack surface along the software supply chain~\cite{cisa_nsa_sbom_2025,woo2021v0finder,deng2025chainfuzz}. Security teams therefore need to know not only that a vulnerability may exist, but also whether it can be triggered in the affected version under realistic runtime conditions, and what its actual security impact is~\cite{householder2020exploit,jacobs2023epss}.

However, a significant gap remains between public vulnerability information and the runtime evidence needed for defensive decisions. The information required to reproduce a CVE is usually scattered across CVE entries, vendor advisories, patch commits, and PoC repositories. These sources describe different aspects, such as the root cause, the affected version, or a partial trigger method~\cite{dong2019inconsistencies}. They rarely provide the complete trigger preconditions and reliable success criteria~\cite{bhandari2021cvefixes,fan2020bigvul,nikitopoulos2021crossvul,mei2024arvo}. Prior studies have also found that public vulnerability databases generally lack directly usable reproduction evidence~\cite{mu2018understanding,ullah2025cvegenie}. Converting fragmented vulnerability disclosures into verifiable runtime evidence is therefore essential to bridging vulnerability intelligence and practical risk assessment.

Historically, end-to-end vulnerability reproduction has depended on costly manual analysis. Mu et al.~\cite{mu2018understanding} coordinated 43 security analysts, who spent more than 3{,}600 hours reproducing only 368 Linux memory vulnerabilities. In recent years, large language model (LLM)-based agents have opened new automation opportunities for code understanding and interactive execution~\cite{fu2023chatgpt,guo2025frontier,ullah2024llms,yao2023react,yang2024sweagent,chang2024aifuzzing,openai2025aardvark,carlini2026zerodays}. In a recent incident, multiple OpenAI models escaped their test environments through combined attacks and reached Hugging Face's production infrastructure~\cite{openai2026hfincident}. DARPA's AI Cyber Challenge also shows that autonomous cyber reasoning systems can perform vulnerability discovery and repair on real open-source software~\cite{zhang2026aixcc}. Recent multi-agent vulnerability exploitation systems further demonstrate that task decomposition and refinement driven by execution feedback can improve performance on complex security tasks~\cite{chen2026vulnsage,zhu2026teams}. For example, CVE-GENIE organizes vulnerability information processing, environment construction, exploit generation, and result validation into an end-to-end pipeline. It successfully reproduced 428 of 841 CVEs~\cite{ullah2025cvegenie}. CVE-Factory converts sparse CVE information into executable task packages that include runtime environments, vulnerability tests, and reference fixes~\cite{luo2026cvefactory}. These efforts verify the feasibility of LLM-based multi-agent systems for processing CVEs at scale.

\begin{figure}[t]
\centering
\includegraphics[width=\columnwidth]{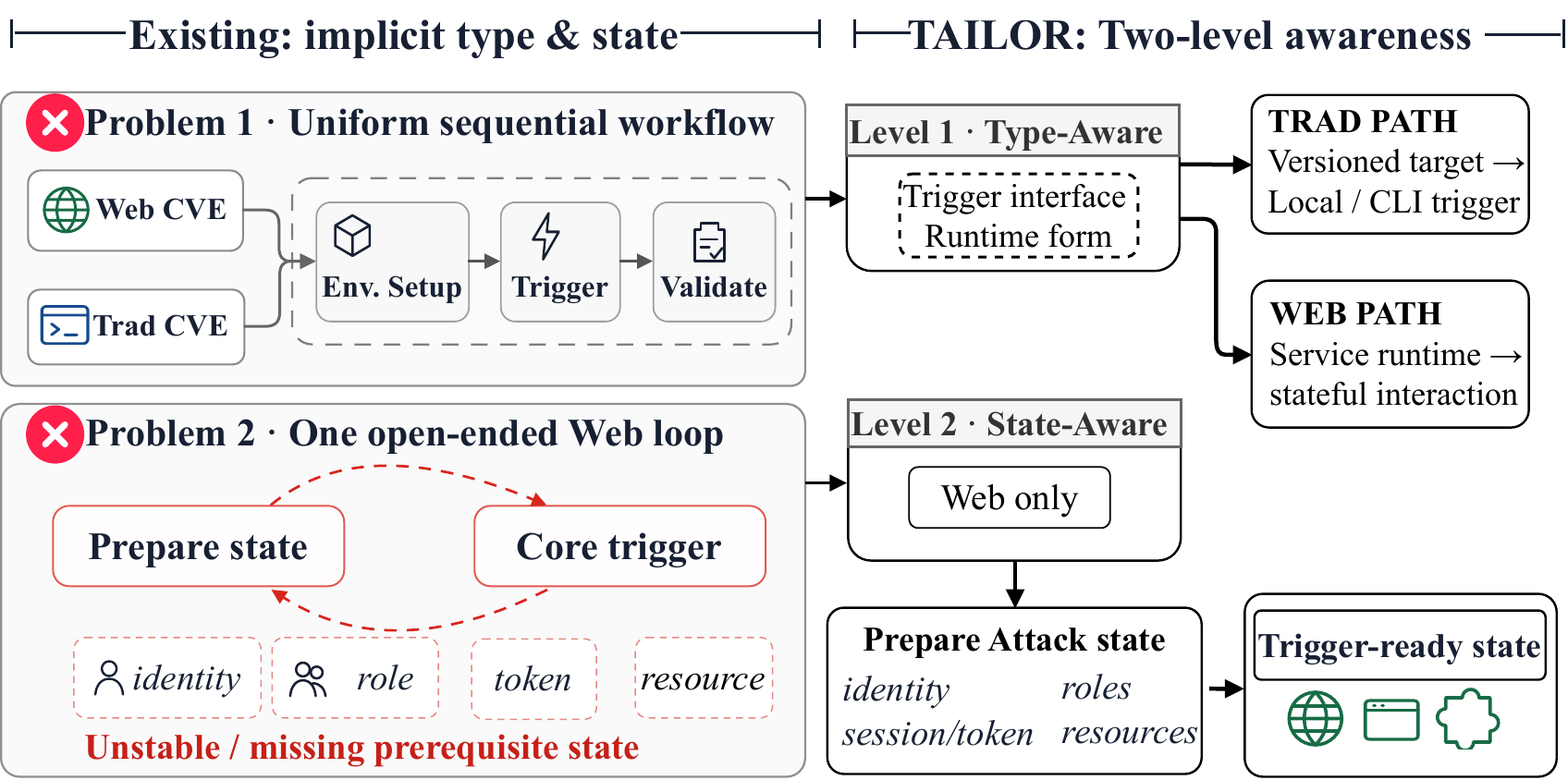}\par
\caption{Limitations of uniform CVE-reproduction workflows and TAILOR's two-level type- and state-aware organization.}
\label{fig:motivation}
\end{figure}

Although existing systems demonstrate that LLM-based multi-agent systems can support large-scale vulnerability reproduction, they usually process all CVEs through a uniform workflow. They leave the reproduction differences among CVEs to be handled implicitly by the agents inside each stage. This design overlooks how vulnerability types differ in their targets' runtime forms and trigger interfaces. These differences further affect the execution workflow, making a uniform process difficult to adapt to their diverse reproduction requirements. For traditional vulnerabilities, reproduction difficulty usually lies in establishing the correct target environment and exposing a latent trigger path; for Web vulnerabilities, it lies in constructing and maintaining distributed application state. The differences among complex vulnerabilities therefore also change environment readiness conditions and cross-stage information dependencies. In particular, when a Web vulnerability's state preparation and core trigger are coupled in the same open-ended agent loop, complex state dependencies are difficult to maintain stably and to pass on to subsequent verification stages. This increases ineffective exploration and weakens the verifiability of the reproduction result. Prior work also shows that complex service configurations and missing authentication information significantly reduce the Web reproduction performance of LLM agents~\cite{liu2025arewethereyet}. Figure~\ref{fig:motivation} summarizes the two architectural limitations of existing uniform workflows and the corresponding TAILOR mechanisms that address them.

To address these problems, we present TAILOR, an end-to-end multi-agent reproduction framework for complex Web CVEs and traditional CVEs, built on type awareness and state awareness. Starting from a CVE identifier, TAILOR automatically converts fragmented CVE metadata into a tailored agent task. TAILOR adopts a two-level control architecture. The first level is a type-aware mechanism. It selects a matching execution path according to the vulnerability's trigger interface and the target's runtime form, so that different reproduction requirements are no longer constrained by a uniform workflow. The second level applies state awareness to the Web path. Before exploitation, it explicitly constructs the minimal application state and the trigger plan required by the vulnerability, and converts them into execution constraints shared by subsequent stages. In this way, TAILOR starts from public vulnerability information and combines the two mechanisms to turn each vulnerability-triggering process into an executable and traceable baseline of security facts, thereby providing a unified and auditable basis for vulnerability confirmation and automated security analysis.

In summary, this work makes the following contributions:
\begin{itemize}
  \item We design and implement TAILOR, an end-to-end multi-agent reproduction framework for complex Web CVEs and traditional CVEs. TAILOR converts fragmented public vulnerability information into independently verified runtime evidence, and packages the reproduction environment, process PoCs, and other behavioral records as an auditable technical basis for vulnerability impact assessment and defense.
  \item We propose a two-level reproduction control method centered on type awareness and state awareness. The type-aware mechanism selects a matching execution path, and the state-aware mechanism decouples the triggering process of complex vulnerabilities into prerequisite state preparation and the core trigger. Together, the two mechanisms let each vulnerability type follow an execution process matched to its reproduction requirements.
  \item We construct an evaluation dataset of 200 CVEs. It covers 160 distinct projects, 16 programming languages, and 57 CWE classes, spanning diverse and complex vulnerability reproduction requirements. It provides a unified basis for evaluating end-to-end automated reproduction of different types of CVEs.
  \item We systematically evaluate TAILOR on this dataset. The results show that TAILOR adapts to the different reproduction requirements of Web and traditional CVEs, and reproduces complex state-dependent cases that a uniform pipeline misses. They further demonstrate that explicitly modeling the differences among CVEs in their execution process and state requirements is an effective way to expand the coverage of automated vulnerability reproduction and produce trustworthy vulnerability instances.
\end{itemize}

\section{Related Work and Limitations}
\label{sec:related}

This section reviews related work on end-to-end CVE reproduction and vulnerability exploitation, and analyzes the limitations of existing methods in reproduction workflow organization and prerequisite Web state handling. Table~\ref{tab:related} compares the main capabilities of representative systems.

\subsection{End to End CVE Reproduction}
\label{sec:related-repro}

To reconstruct runnable, verifiable vulnerability instances from public CVE information, existing studies mainly follow two directions. One directly constructs end-to-end frameworks that cover environment reconstruction, vulnerability triggering, and result validation. The other organizes these reproduction artifacts as executable security tasks.

For end-to-end reproduction, CVE-GENIE divides reproduction into a linear process of information processing, environment construction, vulnerability exploitation, and result validation~\cite{ullah2025cvegenie}. Lotfi et al. use retrieval to supplement missing information and generate containerized environments, attack code, and validation tests from preconditions and postconditions~\cite{lotfi2025automated}. These systems directly connect CVE information to validation results through a complete execution chain. Another line of work uses vulnerability reproduction to construct executable security tasks. CVE-Factory recovers runtime environments, vulnerability tests, and reference fixes through staged generation and progressive validation, converting sparse CVE information into tasks executable by coding agents~\cite{luo2026cvefactory}. Other studies generate PoCs for a given target or reconstruct specific types of vulnerability instances from security patches~\cite{wang2025cybergym,pu2026patchtopoc}. Although their application goals differ, these methods likewise need to recover the environment, trigger conditions, and outcome criteria~\cite{mei2024arvo,bui2022vul4j,bhuiyan2023secbench,dashevskyi2014testrex}.

These methods reduce the execution complexity of reproduction through multistage decomposition~\cite{he2025multiagent,lotfi2025automated,luo2026cvefactory,ullah2025cvegenie,li2025patchpilot}. However, existing frameworks usually use a uniform reproduction process for different CVEs and leave their specific differences to be handled within individual stages. The target runtime form and vulnerability trigger interface further affect environment readiness conditions, subsequent trigger methods, and information dependencies between stages. A uniform process cannot fundamentally address these differences in execution requirements. Existing methods rarely match the process to each CVE before execution, so CVEs with different runtime forms and triggers still share one pipeline.

\subsection{Vulnerability Exploitation}

Vulnerability exploitation determines whether a deployed target can be driven to exhibit the vulnerability and is therefore a key component of end-to-end vulnerability reproduction~\cite{avgerinos2011aeg,cha2012mayhem,wang2018revery,wu2018fuze,alhuzali2018navex}. Existing methods improve this stage through program analysis, runtime feedback, and specialized agent roles~\cite{desai2026pagent,simsek2025pocgen,nitin2025faultline,chen2026vulnsage}. PAGENT combines lightweight rule-based static analysis with sanitizer profiling and coverage feedback to guide a PoC-generation agent~\cite{desai2026pagent}. PoCGen combines vulnerability descriptions with static and dynamic analysis to generate and iteratively repair PoCs~\cite{simsek2025pocgen}. FaultLine organizes hierarchical reasoning around propagation paths from externally accessible inputs to security-sensitive operations~\cite{nitin2025faultline}. Together, these systems use program structure and execution signals to localize the vulnerability and synthesize a trigger.

Recent work also reorganizes tasks within the exploitation stage. Cve2PoC separates attack-strategy replanning from PoC implementation repair into strategic and tactical feedback loops, preventing a failure at one layer from forcing unproductive debugging at the other~\cite{cve2poc2026}. VulnSage assigns vulnerability analysis, exploit generation, execution validation, and failure diagnosis and reflection to specialized agents~\cite{chen2026vulnsage}. These systems make analysis, generation, and refinement more explicit, but they still expose vulnerability triggering to the surrounding reproduction pipeline as one stage rather than separating prerequisite-state construction from core-trigger execution.

For Web vulnerabilities, however, the core trigger often assumes that application state has already been established. Web security research has long recognized the importance of stateful interaction in vulnerability testing~\cite{doupe2012enemy,atlidakis2019restler,lyu2023miner,deng2023nautilus,chen2023synthdb,dahse2014secondorder}. State-aware scanners therefore explore navigation paths or request sequences to reach security-relevant server states. Recent systems make this dependency especially clear. EvoCrawl searches sequences of Web interactions to traverse server-side states and execute code reachable only under particular application states~\cite{guo2025evocrawl}. VOAPI2 derives stateful REST request sequences from producer--consumer dependencies and API semantics, injects vulnerability-oriented payloads, and validates results through feedback~\cite{du2024voapi2}. JAEX manipulates shared objects across multiple requests to expose cross-thread data flows in Java Web applications and guide exploit generation~\cite{huang2025jaex}. These systems show that application state and interaction order are first-class concerns in Web vulnerability testing. Nevertheless, as Table~\ref{tab:related} summarizes, general CVE reproduction frameworks still leave prerequisite-state construction inside the exploitation loop rather than exposing it as a separate stage with an explicit handoff to core-trigger execution~\cite{ullah2025cvegenie,lotfi2025automated,cve2poc2026}.

In Table~\ref{tab:related}, VOAPI2's $\sim$ denotes functionality-conditioned test selection, while EvoCrawl's $\sim$ entries denote pluggable triggering and verification scanners.

\begin{table}[t]
\centering
\caption{Capabilities of representative CVE reproduction and vulnerability exploitation systems. \cmark: primary capability; $\sim$: limited or implicit; \xmark: unsupported in the described workflow.}
\label{tab:related}
\footnotesize
\setlength{\tabcolsep}{3pt}
\begin{tabular}{lcccccc}
\toprule
System & Env. & Trigger & Verify & Route & State & E2E \\
\midrule
CVE-GENIE~\cite{ullah2025cvegenie} & \cmark & \cmark & \cmark & \xmark & $\sim$ & \cmark \\
Automated V\&V~\cite{lotfi2025automated} & \cmark & \cmark & \cmark & \xmark & \xmark & \cmark \\
CVE-Factory~\cite{luo2026cvefactory} & \cmark & \cmark & \cmark & \xmark & \xmark & \cmark \\
Cve2PoC~\cite{cve2poc2026} & \xmark & \cmark & \cmark & \xmark & \xmark & $\sim$ \\
EvoCrawl~\cite{guo2025evocrawl} & \xmark & $\sim$ & $\sim$ & \xmark & \cmark & \xmark \\
VOAPI2~\cite{du2024voapi2} & \xmark & \cmark & \cmark & $\sim$ & \cmark & \xmark \\
JAEX~\cite{huang2025jaex} & \xmark & \cmark & \cmark & \xmark & \cmark & \xmark \\
\midrule
TAILOR & \cmark & \cmark & \cmark & \cmark & \cmark & \cmark \\
\bottomrule
\end{tabular}
\end{table}

\section{Design Rationale}
\label{sec:problem}

\S\ref{sec:related} identifies two limitations of existing CVE reproduction systems. A uniform process hides the reproduction requirements of different CVEs, and prerequisite Web state is not separated from the vulnerability exploitation stage. This section analyzes how these two limitations cause reproduction failures and derives the design goals of TAILOR.

\begin{figure*}[t]
\centering
\includegraphics[width=1\textwidth,trim=0 5bp 0 0,clip]{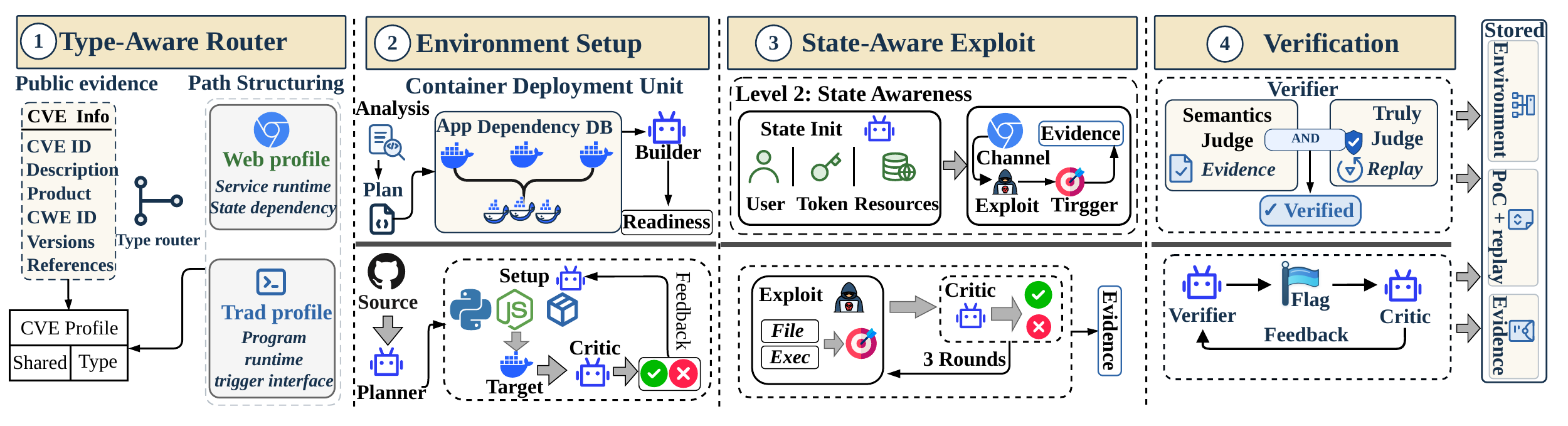}\par
\caption{TAILOR architecture with type-aware routing into a state-aware Web DAG and a traditional sequential pipeline.}
\label{fig:overview}
\end{figure*}

\subsection{Failure Analysis}
\label{sec:challenges}

\noindent\textbf{Uniform Workflow Across CVEs.}
Existing approaches typically use a uniform pipeline to reproduce different types of CVEs. This design assumes that agents can adapt to different reproduction requirements through analysis within each stage. However, differences among complex vulnerability types in runtime form, trigger interface, and related characteristics create distinct dependencies and execution requirements across reproduction stages. A fixed pipeline therefore faces an inherent adaptability tradeoff: optimizing it for one vulnerability type reduces its applicability to others, whereas accommodating multiple types requires compromises in stage design that constrain overall reproduction coverage.

For traditional CVEs, uncertainty usually lies in restoring a version-correct target under build and dependency constraints. Once the target is running, the vulnerability can usually be triggered through a local execution interface, and its effect can be observed within the same execution context. In contrast, successful deployment of a Web vulnerability environment does not mean that the vulnerability is ready to be triggered. The target may still need to satisfy specific application-state constraints~\cite{doupe2012enemy,chen2023synthdb,dahse2014secondorder}, and the trigger may span several interactions~\cite{atlidakis2019restler,lyu2023miner,deng2023nautilus}. Web and traditional CVEs therefore have different execution requirements, making a uniform process difficult to adapt to both types.

\noindent\textbf{Coupled Web Preparation and Triggering.}
Existing frameworks usually treat vulnerability triggering as a single execution stage. They do not further separate it into state preparation and trigger execution. For a Web CVE, the core trigger often depends on application state that is established in advance and reused by later attack steps~\cite{doupe2012enemy,chen2023synthdb,dahse2014secondorder}. For example, an access-control vulnerability may require a legitimate identity to create a resource with an ownership relation before another identity accesses that resource. Existing vulnerability-triggering stages couple state preparation with the specific trigger. State preparation establishes runtime facts that later operations can reuse; triggering should then execute the core trigger against relatively stable state. If both are placed in the same open-ended loop, the agent must maintain two objectives at the same time. Their information and feedback compete for limited context and tool budgets. Prior evaluations also show that complex service configuration and missing authentication information significantly reduce Web vulnerability reproduction performance~\cite{liu2025arewethereyet}. Recent Web agents can use browser interaction, attack planning, and persistent state to complete multistep exploitation~\cite{liu2025arewethereyet,jaswal2026awe,sajadi2026axe}. However, prerequisite state is usually provided by the task context and is not explicitly modeled as an execution stage. Web vulnerability reproduction should therefore separate vulnerability triggering into an attack-preparation stage and a specific vulnerability-triggering stage.

These two types of failure affect each other. Workflow mismatch imposes incorrect stage boundaries, while missing prerequisite state increases ineffective exploration and may even block the intended trigger. The root cause is not simply insufficient agent capability. Type and state requirements remain implicit during execution.

\subsection{Design Goals}
\label{sec:principles}

The preceding analysis leads to two design goals.

\noindent\textbf{Goal 1: Type-aware execution.}
The system should select a matching execution path according to the vulnerability trigger interface and the target runtime form. Web CVEs enter a path that supports service deployment and stateful interaction. Traditional CVEs retain a sequential path for local execution. The two paths define their own stage boundaries and readiness conditions.

\noindent\textbf{Goal 2: State-aware coordination.}
Complex Web vulnerability exploitation should be decomposed into two connected stages: prerequisite-state preparation and core-trigger execution. Before exploitation, the system should identify and construct the application state required by the trigger and expose that state as explicit constraints on exploitation and verification. The Exploit Agent can then reuse the established state and focus on the core vulnerability logic.

These two goals form the two-level design of TAILOR. Type awareness selects an execution path that matches the reproduction requirements. State awareness coordinates attack preparation and vulnerability exploitation in the Web path.

\section{System Design and Implementation}
\label{sec:design}

TAILOR is a multi-agent framework that reproduces Web and traditional CVEs from a CVE identifier and its publicly available vulnerability information. It reconstructs the target environment, triggers the vulnerability, verifies the vulnerability effect against an evidence contract, and preserves the reproduction artifacts required for replay and audit.

TAILOR introduces two levels of control. The first level provides type awareness and routes each CVE to either the Web path or the traditional path. The second level provides state awareness and applies only to the Web path. It treats application state as an explicit element of vulnerability reproduction and constructs the required prerequisite state before exploitation.

Figure~\ref{fig:overview} presents the overall two-level architecture of TAILOR. The Web path is the main design contribution of this work, while the traditional path retains the sequential reproduction logic of CVE-GENIE. Sections~\ref{sec:env} through~\ref{sec:verify} focus on the design of the Web path. Section~\ref{sec:preanalysis} describes the routing and data-structuring process shared by both paths. Section~\ref{sec:env} presents environment planning and reconstruction, with an emphasis on environment construction in the Web path. Section~\ref{sec:webexp} separates Web attack preparation and vulnerability exploitation into two connected stages with distinct responsibilities. Section~\ref{sec:verify} presents evidence-based verification and replay. Section~\ref{sec:artifact-preservation} describes the preservation of reusable reproduction artifacts. Finally, Section~\ref{sec:trad-pipeline} explains how the traditional path adapts the CVE-GENIE workflow within the TAILOR framework. Appendix~\ref{app:implementation} provides implementation and coordination details, and Appendix~\ref{app:cwe-coordination} lists the CWE-indexed Web defaults.

\subsection{Type-aware Routing and Data Structuring}
\label{sec:preanalysis}

TAILOR organizes type-aware routing and data structuring as a sequential pre-analysis process. This process selects a matching execution workflow for each CVE and provides the workflow with actionable vulnerability information.

\begin{figure}[t]
\centering
\includegraphics[width=\columnwidth]{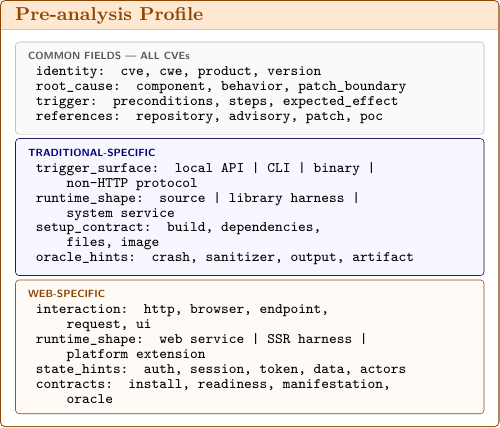}\par
\caption{Path-specific pre-analysis profile.}
\label{fig:preanalysis-schema}
\end{figure}

\noindent\textbf{Type-aware routing.}
\label{sec:routing}
The Routing Layer extracts multiple signals from public vulnerability materials. It focuses on the vulnerability description and patch context, supplemented by affected-version ranges and repository metadata. TAILOR uses the required execution context and interaction medium as the primary basis for routing.

A CVE enters the Web pipeline when its trigger depends on a long-running Web service and its network-facing stack. Typical cases require an HTTP request, browser interaction, or server-side rendering. A CVE enters the traditional pipeline when its trigger surface is a program-level execution interface rather than a Web-service interaction surface.

TAILOR also uses the target software's runtime form as a secondary signal. This signal reduces routing errors caused by product labels alone. For example, an npm package should not enter the traditional pipeline simply because it is distributed as a package. If its vulnerable behavior occurs during server-side rendering, TAILOR routes it to the Web pipeline.

\noindent\textbf{Path-conditioned data structuring.}
After routing, TAILOR structures the CVE's public information for the selected path, as shown in Figure~\ref{fig:preanalysis-schema}. Facts required by both paths form a shared knowledge core, including the scope of impact and environment prerequisites. Execution details required by the selected path are stored in path-specific extensions.

Traditional CVEs usually require a source build before the program is invoked at a trigger entry point. The path-specific extensions for traditional CVEs therefore emphasize \texttt{setup\_\allowbreak contract}, which describes build and dependency preparation, and \texttt{trigger\_\allowbreak surface}, which specifies the program interface required to trigger the vulnerability.

Web CVEs depend on a running service. Their triggers often require a specific interaction mode and preconstructed state. The Web extensions therefore emphasize \texttt{interaction}, which describes how to interact with the service, and \texttt{state\_\allowbreak hints}, which identifies the state that must exist before triggering.

This split avoids duplicating common fields across paths while preserving the execution-relevant differences.

\subsection{Environment Planning and Reconstruction}
\label{sec:env}

\begin{figure}[!b]
\centering
\includegraphics[width=\columnwidth]{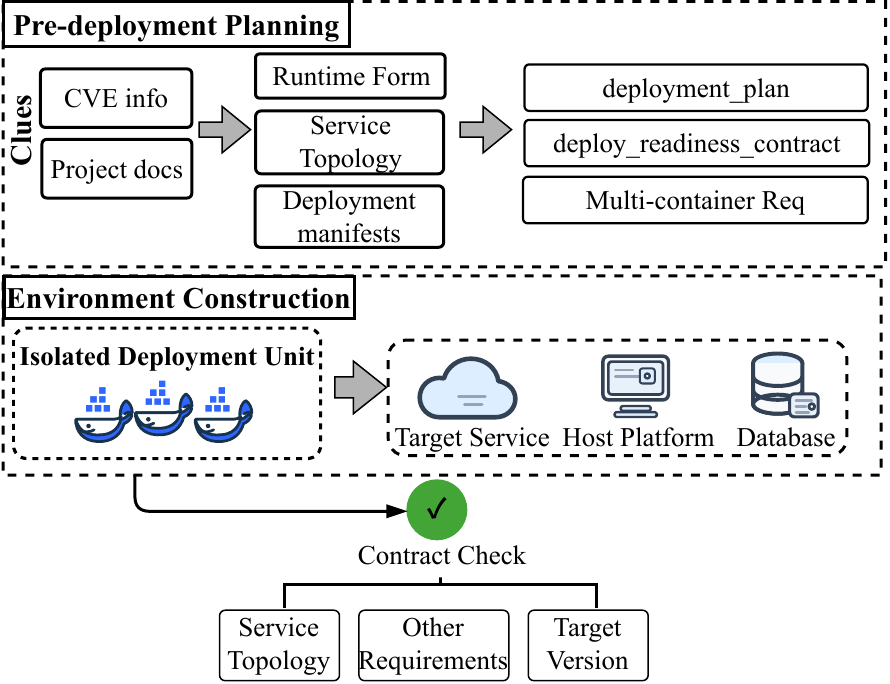}\par
\caption{Detailed environment planning, construction, and readiness validation.}
\label{fig:environment-detail}
\end{figure}

After routing, TAILOR turns the CVE configuration for the selected path into an isolated target runtime environment. Public vulnerability materials usually identify the affected version, but they lack the service topology and readiness conditions required to run the target~\cite{mei2024arvo,dashevskyi2014testrex}. To fill this gap, TAILOR performs deployment planning, environment construction, and readiness validation in sequence. Figure~\ref{fig:environment-detail} presents this three-step process in detail.

\noindent\textbf{Pre-deployment planning.}
TAILOR parses the project documentation and deployment manifests to determine the target version, runtime form, and service topology. This\nopagebreak[4] stage generates two structured artifacts. \texttt{deployment\_\allowbreak plan}\nopagebreak[4] describes how the environment is constructed. \texttt{deploy\_\nopagebreak[4]\allowbreak readiness\_contract} defines the conditions that the environment must satisfy before it enters the attack preparation stage.

\noindent\textbf{Environment construction.}
Each CVE corresponds to an isolated container deployment unit. The unit can range from a single target container to a multi-container topology that includes a database and other external dependencies. TAILOR prefers deterministic orchestration. When a usable container image and sufficient planning information are available, the \emph{WebEnvBuilder} agent completes the deployment inside the orchestrated unit according to \texttt{deployment\_plan}. Otherwise, the agent autonomously analyzes and deploys the container unit. It completes the remaining construction according to \texttt{deployment\_plan} and \texttt{deploy\_readiness\_contract}.

\noindent\textbf{Readiness validation.}
\looseness=-1
To curb agent hallucination, TAILOR applies two layers of checks to every constructed unit. It first performs basic health checks to confirm that the container is alive and its port is reachable. The \emph{WebEnvCritic} then performs a contract audit according to \texttt{deploy\_readiness\_contract}. The unit can proceed to the next stage only when the target version, service topology, and the remaining contract clauses are all satisfied. When validation fails, TAILOR summarizes the verified facts, completed steps, failed attempts, blockers, and next action into an experience record. It injects the record into the next agent iteration, so that subsequent attempts can reuse completed progress and avoid repeating failed steps~\cite{shinn2023reflexion}.

\subsection{State-Aware Web Attack Preparation and Exploitation}
\label{sec:webexp}

\begin{figure}[H]
\centering
\includegraphics[width=\columnwidth]{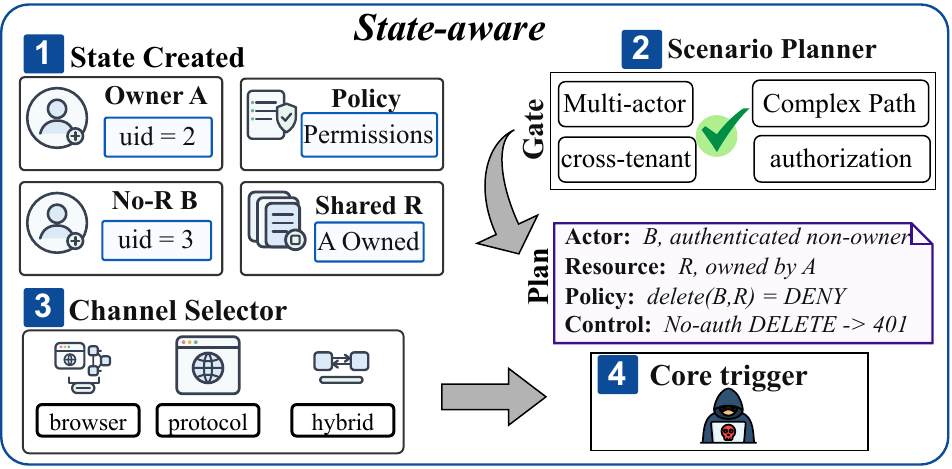}\par
\caption{State-aware attack preparation with scenario planning and channel selection for multi-actor Web vulnerabilities.}
\label{fig:state-aware-preparation}
\end{figure}

Reproducing complex Web CVEs requires constructing and maintaining a vulnerability trigger chain. Its trigger conditions are distributed across multiple state domains rather than concentrated in a single request. These domains include client sessions, server-side business state, accounts with specific privileges, and dynamic tokens. If an Exploit Agent interleaves state construction with vulnerability triggering, it must maintain two conflicting objectives at the same time. State construction modifies the application state, while triggering requires a stable state. TAILOR addresses this problem by introducing a second-level state-aware mechanism. It decouples exploitation into attack preparation (\S\ref{sec:state-aware-prep}) and vulnerability execution (\S\ref{sec:exploit-exec}). As shown in Figure~\ref{fig:state-aware-preparation}, the attack-preparation stage converts the deployment-ready environment into a validated trigger-ready state. It then optionally generates a scenario contract and selects an attack channel. The execution stage focuses on the core vulnerability logic.

\subsubsection{State-Aware Attack Preparation}
\label{sec:state-aware-prep}

In this paper, we define state as runtime facts in the target application that are observable and affect subsequent trigger outcomes. These facts include accounts and privileges, verified credentials, sessions, and tokens. We implement the state-aware attack preparation stage through the following three sequential steps.

\noindent\textbf{State initialization.}\label{sec:stateful}
An LLM first derives the minimal prerequisite state from the CVE's trigger preconditions and the current application state. It then combines the CWE class with this result to determine the final authentication and role requirements. The \emph{StateInitializer} agent constructs this state in the target environment and produces a trigger-ready configuration. This removes the prerequisite reasoning burden from the downstream Exploit Agent.

\noindent\textbf{Scenario contract.}\label{sec:scenario-gate}\label{sec:scenario-plan}
A lightweight scenario gate first determines whether the vulnerability depends on multiple roles or business states. Vulnerabilities that do not depend on them skip scenario planning. For dependent vulnerabilities, the \emph{Scenario Planner} generates a contract over roles, resources, and policies based on \texttt{state\_result}. The contract specifies the positive evidence and negative controls required for subsequent validation.

\noindent\textbf{Attack channel selection.}\label{sec:channel}
TAILOR selects a channel according to the vulnerability's interaction surface. The protocol channel is used for server-side vulnerabilities that can be precisely triggered through raw HTTP requests. The browser channel is used for vulnerabilities that depend on DOM manipulation, JavaScript events, or page context. The hybrid channel is used when established browser or authentication state is combined with protocol-level requests.

\subsubsection{Web Exploitation}
\label{sec:exploit-exec}

\noindent\textbf{Input.}
The \emph{Web Exploit Agent} receives the trigger-ready environment and \texttt{scenario\_result}, together with the trigger specification in Figure~\ref{fig:preanalysis-schema}. The specification covers three field groups: \texttt{trigger} (\texttt{preconditions}, \texttt{steps}, and \texttt{expected\_effect}), \texttt{interaction} (\texttt{http}, \texttt{browser}, \texttt{endpoint}, \texttt{request}, and \texttt{ui}), and \texttt{references.poc}.

\noindent\textbf{Exploit agent.}
Within a bounded ReAct loop~\cite{yao2023react}, the agent constructs attack requests, reuses the prerequisite state in \texttt{state\_result}, and executes the trigger subject to the constraints in \texttt{scenario\_result}. For vulnerabilities that require differential comparison, it executes the baseline and attack steps and preserves the pre-attack and post-attack states required for validation. An error detector continuously monitors the loop. When it identifies repeated failures of the same class, the execution controller terminates the current round. The controller summarizes the verified facts, completed operations, and failed paths into a structured experience record and injects it into the next attempt to avoid wasting budget~\cite{shinn2023reflexion}.

After a successful trigger, the agent invokes {\footnotesize\texttt{save\_evidence}} to record the PoC, key trigger operations, the tool-call trace, and the raw runtime evidence. The normalized results are used for subsequent verification.

\subsection{Verification}
\label{sec:verify}

\begin{figure}[t]
\centering
\includegraphics[width=\columnwidth]{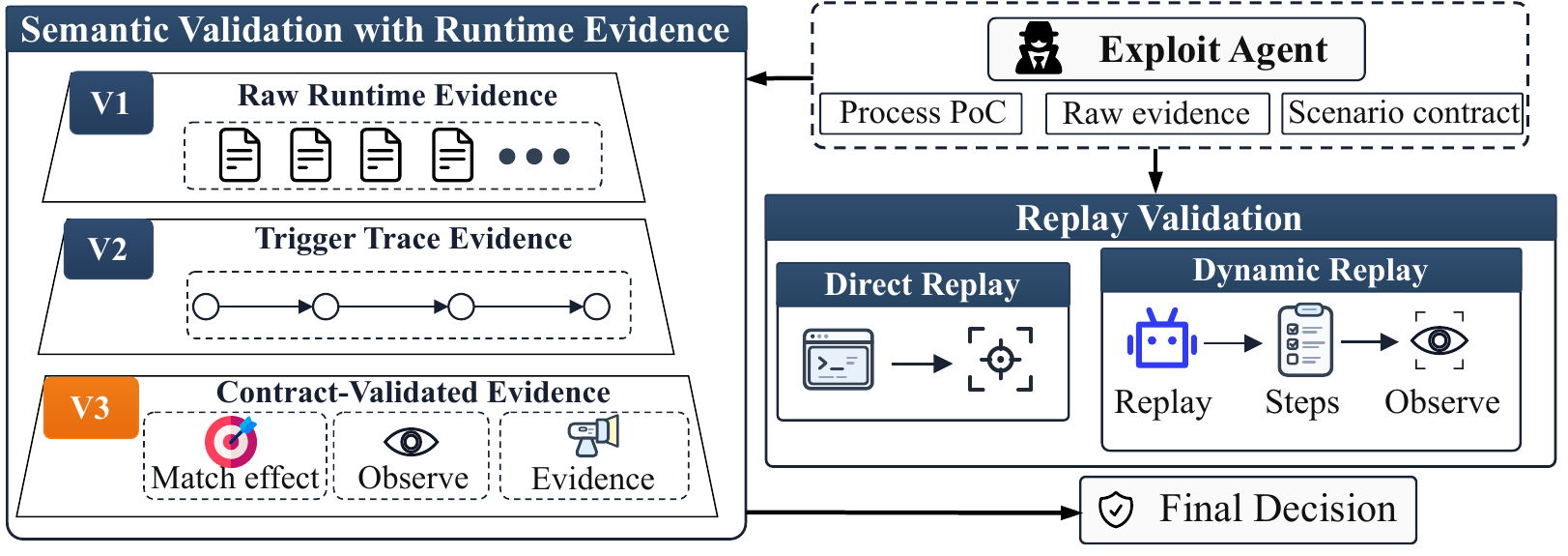}\par
\caption{Detailed verification process combining semantic validation, replay, and final decision.}
\label{fig:verification-detail}
\end{figure}

After exploitation ends, TAILOR does not directly trust the Exploit Agent's success report~\cite{lin2025llm,zhang2025mirage}. For a Web CVE, even a successful request--response exchange shows only that an interaction occurred. It does not establish that the target security boundary was violated. TAILOR therefore uses an independent Verifier that combines evidence review with constrained replay, as shown in Figure~\ref{fig:verification-detail}.

\subsubsection{Semantic Validation with Runtime Evidence}

The Verifier performs a three-level semantic validation of the available evidence. Each level imposes stronger constraints on the output of the preceding level.

\textbf{V1 (raw runtime evidence)} records the outputs returned directly by tools during vulnerability reproduction. Together, these records form an auditable execution trace. \textbf{V2 (trigger-trace evidence)} uses evidence provenance and temporal order to organize the V1 records into a candidate trigger chain. This chain links the triggering actions to their runtime observations. \textbf{V3 (contract-constrained evidence)} checks each element of the V2 trace against the success contract for the target CVE. The observed result must match the expected vulnerability effect, and the effect must be visible at a direct observation point. Evidence unrelated to the core vulnerability semantics is excluded. Only a reproduction that passes all three levels is accepted.

Vulnerability classes produce different observable effects. The Verifier therefore selects type-specific validation signals and benign controls. Table~\ref{tab:cwe-templates} summarizes the observation criteria for representative vulnerabilities and the controls used to rule out normal behavior.

\begin{table*}[t]
\centering
\footnotesize
\caption{Verification criteria and benign controls.}
\label{tab:cwe-templates}
\setlength{\tabcolsep}{4pt}
\begin{tabular*}{\textwidth}{@{\extracolsep{\fill}}p{0.17\textwidth}p{0.18\textwidth}p{0.34\textwidth}p{0.23\textwidth}@{}}
\toprule
\textbf{Effect} & \textbf{CWEs} & \textbf{Success evidence} & \textbf{Benign control} \\
\midrule
Code execution
  & {\raggedright CWE-78, CWE-94\par}
  & Unique marker in output, a file, or process state
  & Marker is not a static echo or normal output \\
\addlinespace
Unauthorized access or change
  & {\raggedright CWE-284, CWE-639, CWE-863\par}
  & Attacker reads or changes a victim resource
  & A policy-consistent baseline denies the action \\
\addlinespace
Browser script execution
  & {\raggedright CWE-79\par}
  & DOM change, alert or console marker, or network canary
  & Payload appears only in response text or page source \\
\addlinespace
Unauthorized server-state change
  & {\raggedright CWE-352\par}
  & Protected server value changes after the attack
  & State predates the attack or changes through an authorized action \\
\bottomrule
\end{tabular*}
\end{table*}

\subsubsection{Validation through Replay}

Semantic validation establishes that the available evidence is consistent with the target vulnerability, but it cannot rule out fabricated evidence. To assess authenticity and repeatability, TAILOR constructs a constrained replay from the process PoC, saved evidence, and scenario contract~\cite{dashevskyi2014testrex,allen2024webrr}. The replay checks whether the same vulnerability effect can be observed again.

\noindent\textbf{Direct replay.} The Verifier first checks whether the PoC is well formed and complete. It then executes the PoC in the target environment. If the PoC contains a formatting defect or an unresolved placeholder, the system allows a bounded number of repair iterations. Direct replay is suitable for vulnerabilities with short execution paths and self-contained PoCs.

\noindent\textbf{Dynamic replay.} State-coupled Web vulnerabilities cannot be replayed through a single PoC execution. When the attack depends on a multi-step workflow, the \emph{Active Verification Agent} uses the existing evidence and scenario contract to reconstruct the workflow as an ordered sequence of steps. It executes these steps, collects new evidence, and returns the evidence to the same Verifier for semantic validation~\cite{allen2024webrr}. This tiered design keeps verification overhead low for short-path vulnerabilities while preserving coverage of stateful Web CVEs.

\subsubsection{Final Decision}

The Verifier combines semantic validation with the independent replay result. A reproduction is accepted when the available evidence reaches V3, or when replay produces new evidence that satisfies the proof contract and is accepted as V3. The decision must satisfy three semantic conditions. \textbf{(i) Effect consistency:} execution that reaches the intended trigger point must produce the expected effect. \textbf{(ii) Direct observability:} the evidence must come from a location where the vulnerability effect can be observed directly. \textbf{(iii) Differential control:} when required by the proof contract, the evidence must include a security-control comparison that rules out normal behavior.

\subsection{Reproduction Artifacts}
\label{sec:artifact-preservation}

After the Verifier confirms a successful reproduction, TAILOR exports the current runtime environment and the process PoC. For failed or inconclusive runs, it retains only process files and structured outputs from completed stages.

\begin{figure}[t]
\centering
\resizebox{0.96\columnwidth}{!}{%
\begin{tikzpicture}[
    font=\scriptsize,
    artifact/.style={
        draw,
        rounded corners=2pt,
        align=left,
        text width=6.15cm,
        minimum height=0.82cm,
        line width=0.65pt,
        inner xsep=5pt,
        inner ysep=3pt
    }
]

\node[artifact, fill=blue!6] (evidence) {
    \textbf{Evidence \& verdict}\\[-0.2mm]
    \texttt{evidence.json} $\cdot$ raw output $\cdot$ verification and verifier records
};

\node[artifact, fill=green!7, below=0.20cm of evidence] (poc) {
    \textbf{PoC \& replay contract}\\[-0.2mm]
    executable/text PoC $\cdot$ \texttt{poc\_manifest}\\replay entry and oracle
};

\node[artifact, fill=orange!9, below=0.20cm of poc] (environment) {
    \textbf{Environment \& state}\\[-0.2mm]
    Compose/source/image $\cdot$ named-volume snapshots\\restore script
};

\node[artifact, fill=black!4, below=0.20cm of environment] (provenance) {
    \textbf{Provenance \& guidance}\\[-0.2mm]
    build/exploit results $\cdot$ manifest $\cdot$ README\\portability warnings
};

\node[
    draw,
    rounded corners=3pt,
    fit=(evidence)(poc)(environment)(provenance),
    inner xsep=6pt,
    inner ysep=7pt,
    line width=0.8pt,
    label={[font=\bfseries\small]above:Preserved reproduction artifacts}
] (bundle) {};

\end{tikzpicture}%
}\par
\caption{Reproduction artifact package exported by TAILOR.}
\label{fig:artifact-bundle}
\end{figure}
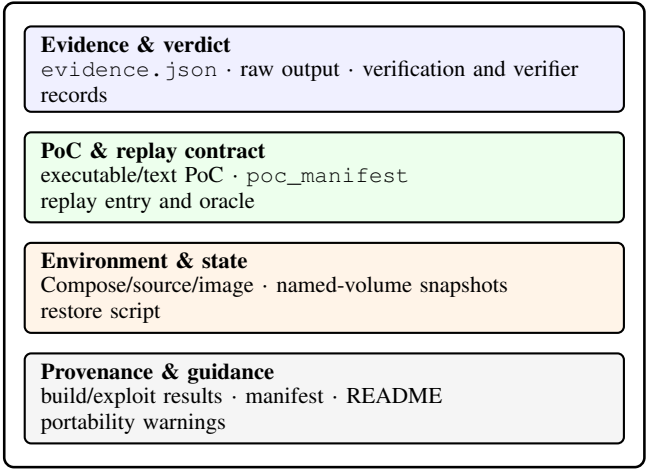

Figure~\ref{fig:artifact-bundle} summarizes the exported package. \emph{Evidence and verdict artifacts} explain why the run was accepted. \emph{PoC and replay artifacts} provide the execution entry point and expected outcome. \emph{Environment and state artifacts} preserve the runtime conditions required to trigger the vulnerability~\cite{mei2024arvo,dashevskyi2014testrex}. \emph{Provenance and guidance artifacts} record the source, use, and portability limits of each item. Together, these materials describe the reproduction scenario~\cite{mei2024arvo,bui2022vul4j,bhuiyan2023secbench,dashevskyi2014testrex}.

TAILOR uses a tiered strategy to preserve the environment. It first retains an available Compose project to record dependencies among services. For a public image, the system pins an immutable digest. When the original project is unavailable, it attempts to preserve the source code or an explicitly configured image. For a deployment with several containers and persistent state, the exporter also attempts to snapshot eligible named volumes. It generates a restoration script for successfully captured state. If only an environment reference or partial artifacts are available, the system still retains the evidence and manifest. It marks the package as \emph{metadata only} or \emph{portability unverified}. It does not claim that the environment can be restored directly on another machine.

\subsection{Traditional Pipeline}
\label{sec:trad-pipeline}
\label{sec:pipeline}

\looseness=-1
Our traditional path builds on the implementation of CVE-GENIE~\cite{ullah2025cvegenie}. For a CVE routed to this path by \S\ref{sec:routing}, TAILOR retains its sequential execution structure. It constructs vulnerability knowledge, analyzes environment prerequisites, builds the affected environment, generates an exploit, and performs CTF-style verification. Environment construction and vulnerability triggering continue to use feedback between developer and critic agents.

Our main changes concern the target environment and the runtime boundary. Before environment construction, the system uses the pre-analysis result to identify the project form. It distinguishes a library target invoked through a local interface from a service target that must be started before access. TAILOR creates a dedicated target container for each traditional CVE to isolate different reproductions. The agent then deploys the environment in this container and normalizes the deployment result into a contract. The contract records the target container's environment information and trigger requirements. The \emph{Exploiter} follows this contract to enter the target environment and trigger the vulnerability.

\looseness=-1
As in the Web path, a successful traditional trigger must preserve its evidence; the run then enters the artifact preservation stage described in \S\ref{sec:artifact-preservation}. TAILOR also uses checkpoints to record completed stages and their outputs. If an interrupted reproduction resumes, it continues from the most recent valid stage. Completed stages do not need to run again. In this way, TAILOR retains the traditional CVE reproduction logic of CVE-GENIE while integrating it with type-aware routing, isolated execution, checkpoint-based recovery, and artifact export.

\section{Evaluation}
\label{sec:eval}

We organize the evaluation around five research questions to examine the reliability and effects of TAILOR's architecture:

\begin{itemize}[leftmargin=1.6em,itemsep=2pt,topsep=3pt,label=\textbullet]
\item \textbf{RQ1.} How effective is TAILOR in end-to-end CVE reproduction?
\item \textbf{RQ2.} Can type-aware routing reliably select the appropriate reproduction path?
\item \textbf{RQ3.} How does type-aware routing affect TAILOR's end-to-end reproduction effectiveness?
\item \textbf{RQ4.} How effective is Level~2 state awareness for complex Web CVE reproduction?
\item \textbf{RQ5.} How does Level~2 state awareness affect TAILOR's reproduction effectiveness on complex Web CVEs?
\end{itemize}

\subsection{Experimental Setup}
\label{sec:eval-setup}

\noindent\textbf{Dataset.} We evaluate TAILOR on a dataset of 200 CVEs. Source~A contains 89 cases from the dataset used by CVE-GENIE~\cite{ullah2025cvegenie}, including 58 Web CVEs and 31 traditional CVEs (i.e., non-Web CVEs). Source~B consists of 111 cases that we independently collected from public CVE records and security advisories (99 Web CVEs and 12 traditional CVEs). When constructing Source~B, we prioritized vulnerabilities that affect Web components in open-source projects. We retained only cases for which the affected version could be identified and the corresponding source code or version archive was publicly available. A public PoC was not an inclusion criterion. We removed cases that overlapped with Source~A based on their CVE IDs. The complete dataset contains 157 Web CVEs and 43 traditional CVEs; Appendix~\ref{app:dataset-composition} gives the selection criteria and composition audit.

\noindent\textbf{Metrics.} RQ1 uses the \emph{Reproduction Success Rate} (RSR), defined as the proportion of CVEs independently verified as \texttt{reproduced}. We report RSR by path, disclosure year, data source, severity, NVD-linked PoC signal, and CWE. Each CVE is executed at most $K=3$ times and contributes one final binary outcome. RQ2 uses routing accuracy and consistency over five runs. RQ3 compares end-to-end RSR with and without type-aware routing. For Web cases with retained or reconstructed parseable attack artifacts, RQ4 measures the number and RSR of cases with explicit state dependencies and their distribution across reproduction outcomes. RQ5 compares end-to-end RSR with and without Level~2 state awareness. We also analyze the cost per case and the budget exceedance rate.

\noindent\textbf{Implementation.} The routing classifier uses Claude Haiku 4.5. The agents responsible for environment construction, vulnerability exploitation, scenario planning, and verifier script generation use Claude Sonnet 4.6 as their base model. The cost limit for each CVE is \$10 on the Web path and \$6 on the traditional path. All experiments are orchestrated by a host machine and run in Docker containers. The models use their default sampling parameters. The complete agent configuration and model assignment are provided in Appendix~\ref{sec:roster}.

\noindent\textbf{Artifact Availability.} The dataset, full reproduction logs, and representative case studies are available at \url{https://anonymous.4open.science/r/TAILOR-EDE2}. The complete implementation will be publicly released upon acceptance.

\subsection{Effectiveness and Generalization}
\label{sec:eval-main}

RQ1 examines TAILOR's end-to-end reproduction effectiveness across diverse CVEs and evaluation conditions.

\noindent\textbf{Overall end-to-end effectiveness.}
Table~\ref{tab:rsr-main} summarizes the end-to-end results. TAILOR reproduces 112 of 200 CVEs, yielding an overall RSR of \textbf{56.0\%}. The Web and traditional subsets reach \textbf{59.2\%} (93/157) and \textbf{44.2\%} (19/43), respectively. Both paths maintain a stable reproduction capability.

\begin{table}[t]
\centering
\footnotesize
\caption{End-to-end reproduction success rate on the full 200-CVE corpus.}
\label{tab:rsr-main}
\renewcommand{\arraystretch}{1.15}
\begin{tabular*}{\columnwidth}{@{\extracolsep{\fill}}lcc@{}}
\toprule
\textbf{Pipeline} & \textbf{Succ./N} & \textbf{RSR} \\
\midrule
Overall & 112/200 & 56.0\% \\
\quad Web & 93/157 & 59.2\% \\
\quad Traditional & 19/43 & 44.2\% \\
\bottomrule
\end{tabular*}
\end{table}

\begin{figure}[t]
  \centering
  \includegraphics[width=0.93\columnwidth]{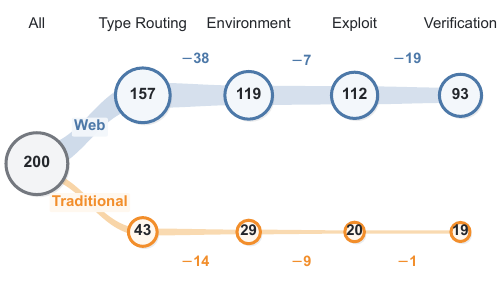}
  \caption{Failure-stage attrition shows where Web and traditional reproductions stop in the end-to-end workflow.}
  \label{fig:failure-attrition}
\end{figure}

\noindent\textbf{Failure stages.}\label{sec:eval-stage} As Figure~\ref{fig:failure-attrition} shows, failures are concentrated in environment construction. Of the 88 failed cases, \textbf{59.1\%} (52/88) stop during environment construction, \textbf{18.2\%} (16/88) stop during exploitation, and \textbf{22.7\%} (20/88) reach verification but do not pass. Thus, \textbf{77.3\%} (68/88) of failures occur before verification. Building an executable environment and producing an effective vulnerability trigger therefore remain the main bottlenecks in end-to-end reproduction.

\begin{figure*}[t]
\centering
\duetbeginsubfigures
\begin{minipage}[t]{0.33\textwidth}
\centering
\includegraphics[width=\linewidth]{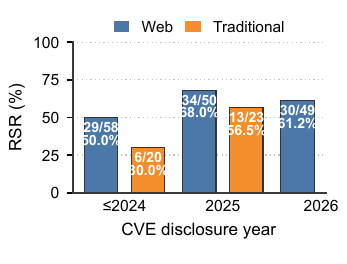}
\duetsubcaption{Disclosure-year robustness.}
\label{fig:rq1-year}
\end{minipage}\hfill
\begin{minipage}[t]{0.27\textwidth}
\centering
\includegraphics[width=\linewidth]{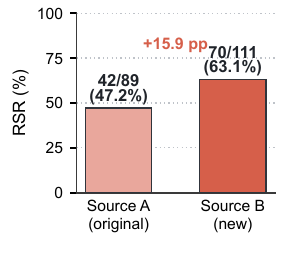}
\duetsubcaption{Generalization across data sources.}
\label{fig:rq1-source}
\end{minipage}\hfill
\begin{minipage}[t]{0.36\textwidth}
\centering
\includegraphics[width=\linewidth]{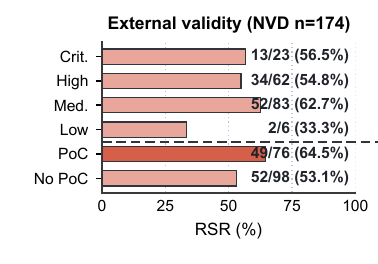}
\duetsubcaption{External-validity splits on the NVD-resolved subset.}
\label{fig:rq1-extval}
\end{minipage}
\duetfinishsubfigures
\caption{RQ1 generalization checks by disclosure year, data source, and NVD-resolved CVSS/PoC splits. Bars show RSR and per-bar counts.}
\label{fig:rq1gen}
\end{figure*}

\begin{figure}[t]
\centering
\includegraphics[width=\linewidth,trim=0 6pt 0 0,clip]{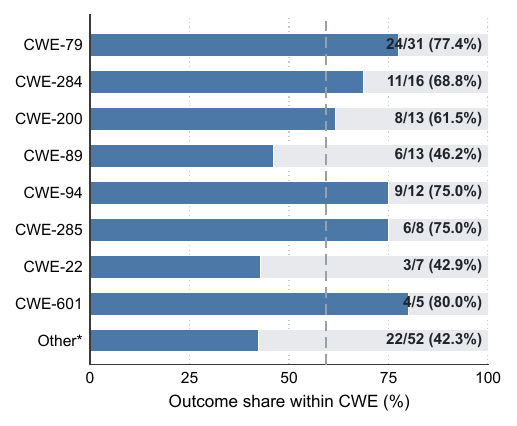}\par
\vspace{-0.45em}
\caption{Per-CWE Web RSR compares reproduction effectiveness across vulnerability families.}
\label{fig:per-cwe}
\end{figure}

\begin{figure}[t]
\centering
\includegraphics[width=0.92\linewidth]{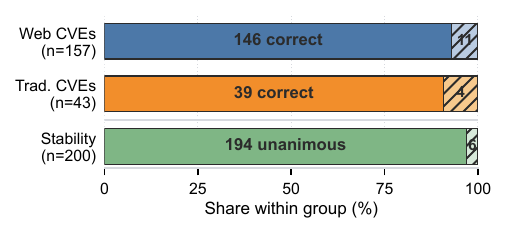}
\caption{Routing reliability summarizes type-aware routing accuracy and stability.}
\label{fig:routing-stability}
\end{figure}

\noindent\textbf{Performance across data sources.}
The two data sources probe different aspects: reproduction on a prior benchmark (Source~A) and generalization to independently collected cases (Source~B). Source~A evaluates TAILOR on a prior benchmark and its failure subset: CVE-GENIE failed to reproduce \textbf{79/89} cases in this source, whereas TAILOR reproduces \textbf{42/89} cases, including \textbf{34/79} of the prior failures. Source~B is an independently collected purpose-oriented case set, on which TAILOR reproduces \textbf{70/111} cases. This result shows that TAILOR's reproduction capability extends to cases that do not overlap with the existing evaluation set (Figure~\ref{fig:rq1gen}(b)).

\noindent\textbf{Performance across data conditions.}
The disclosure-year results show that performance remains stable after the training cutoff. The RSRs for CVEs disclosed in 2024 or earlier, in 2025, and in 2026 are \textbf{44.9\%} (35/78), \textbf{64.4\%} (47/73), and \textbf{61.2\%} (30/49), respectively (Figure~\ref{fig:rq1gen}(a)). Among the 174 cases with NVD CVSS records, the Critical, High, and Medium groups achieve RSRs in a comparable range of \textbf{54.8\%}--\textbf{62.7\%}. In the same subset, the RSR is \textbf{64.5\%} (49/76) when an NVD-linked PoC is present and \textbf{53.1\%} (52/98) when one is absent (Figure~\ref{fig:rq1gen}(c)), indicating that the presence of a PoC signal is associated with a higher RSR, while cases without a signal are still reproduced successfully. Overall, TAILOR performs stably across different evaluation criteria.

\noindent\textbf{Performance across vulnerability classes.}
TAILOR maintains high RSRs for vulnerability classes that rely on directly observable evidence and for those that rely on state-change evidence (Figure~\ref{fig:per-cwe}). The RSR for CWE-79 is \textbf{77.4\%} (24/31). The access-control-related CWE-284 and CWE-285 reach \textbf{68.8\%} (11/16) and \textbf{75.0\%} (6/8), respectively. Successful cases thus span both evidence forms.

\begin{takeaways}
\textbf{Answer to RQ1.}
\textit{TAILOR achieves an end-to-end RSR of \textbf{56.0\%} on 200 CVEs. Its effectiveness is not limited to a single path, data source, or evidence form.}
\end{takeaways}

\label{sec:eval-cwe}

\subsection{Type-Aware Routing Reliability}
\label{sec:eval-paths}

RQ2 evaluates whether type-aware routing makes accurate and repeatable decisions. We run the classifier five times for each of the 200 CVEs, take the majority vote as the routing decision, compare it with the Web and traditional labels used by the evaluation dataset, and measure whether repeated classifications remain stable.

Figure~\ref{fig:routing-stability} shows 146 correctly routed Web cases and 39 correctly routed traditional cases, for \textbf{185/200} correct decisions (92.5\%). Eleven Web cases are sent to the traditional path and four traditional cases to the Web path. Meanwhile, \textbf{194/200} cases (97.0\%) receive the same label in all five runs, with a flip rate of 3.0\% and Fleiss' $\kappa$ of \textbf{0.957}~\cite{fleiss1971agreement}. All 18 decisions made by deterministic rules are unanimous; the LLM subset is unanimous on 176/182 cases ($\kappa=0.951$). Most errors are boundary cases on which the classifier is consistently wrong across runs; the mistakes are systematic rather than random, which further attests to decision stability.

\begin{takeaways}
\textbf{Answer to RQ2.}
\textit{Type-aware routing selects the correct path for \textbf{185/200} CVEs (92.5\%) and gives a unanimous label in all five runs for \textbf{194/200} CVEs (97.0\%), with Fleiss' $\kappa=\textbf{0.957}$.}
\end{takeaways}

\subsection{Effect of Type-Aware Routing}
\label{sec:eval-path-effect}

RQ3 examines the effect of type-aware routing on reproduction. We sample 10 Web CVEs that TAILOR successfully reproduces and, holding the CVE input fixed, force each case through the \texttt{legacy} path to create the No-L1 condition. This measures whether successful cases remain reproducible on the wrong path and thus tests the importance of matching the reproduction path to the CVE. Appendix~\ref{app:ablation-cases} lists the per-case outcomes.

\begin{figure}[t]
\centering
\duetbeginsubfigures
\includegraphics[width=0.92\columnwidth]{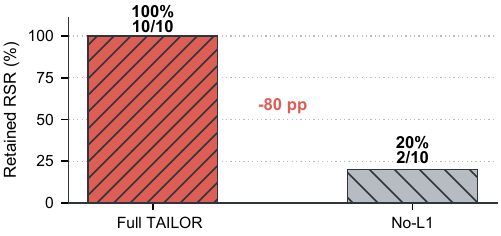}
\duetsubcaption{Success retention.}
\label{fig:no-l1-retention}
\vspace{0.1em}
\includegraphics[width=0.92\columnwidth]{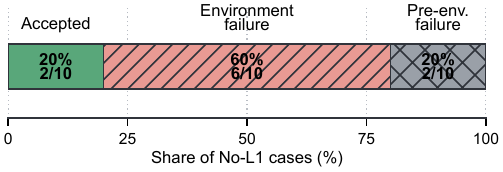}
\duetsubcaption{Terminal outcomes.}
\label{fig:no-l1-outcomes}
\duetfinishsubfigures
\caption{No-L1 evaluates the effect of type-aware routing on success retention and terminal outcomes.}
\label{fig:no-l1}
\vspace{-0.5em}
\end{figure}

Figure~\ref{fig:no-l1-retention} shows that only \textbf{2/10} cases are accepted after being forced onto the traditional path, with the success-retention rate falling to 20\%; the other eight cases all terminate before exploitation (two before environment construction and six during environment construction). Path matching therefore affects the upstream information processing and environment preparation as well as the downstream attack channel: seven of these eight failed cases use a browser or hybrid channel under full TAILOR, reflecting their Web-specific execution requirements. Both accepted cases are converted to local command oracles, and the No-L1 run for CVE-2025-25300 does not reproduce the \texttt{window.opener} browser effect that the full Web path validates. Therefore, \textbf{2/10} is the success-retention rate achieved after further verification.

\begin{takeaways}
\textbf{Answer to RQ3.}
\textit{Among these 10 Web cases reproduced by the full system, removing type-aware routing and forcing a uniform linear workflow retains at most \textbf{20\%}. Type-aware routing is therefore not merely a classification module. It enables a path-specific architecture that matches the requirements of CVE reproduction.}
\end{takeaways}
\vspace{-0.55em}
\subsection{Level 2 State Awareness}
\label{sec:eval-stateful}
\vspace{-0.45em}

RQ4 examines whether Level~2 state awareness can identify the state dependencies of complex Web CVEs and convert them into constraints used during exploitation and verification. We analyze 112 of the 157 Web CVEs with available attack-vector records, first examining the distribution of state dependencies in these cases and then whether TAILOR scales state preparation and scenario planning with the complexity of the state relations. Section~\ref{sec:eval-state-effect} evaluates the effect of Level~2 on reproduction outcomes through the No-L2 experiment.

\noindent\textbf{State dependencies.} We label a case as explicitly state-dependent if its retained attack artifacts record concrete state requirements. Under this criterion, \textbf{93/112} cases have explicit state dependencies, and TAILOR successfully reproduces \textbf{79} of them (RSR \textbf{84.9\%}), indicating that state conditions are a common prerequisite for triggering the majority of complex Web CVEs in this subset. However, the required state handling varies across cases: some require only basic state preparation, while others also require actor and resource relations to be modeled. After identifying a state dependency, Level~2 must determine whether additional scenario planning is required.

\noindent\textbf{On-demand planning.} Among the 112 cases, 86 retain scenario-gate results, and 40 enter scenario planning, covering all 19 multi-actor cases; none of the remaining 46 cases involves multiple actors. The scenario gate therefore identifies cases that require additional relation modeling while sparing the other cases unnecessary planning overhead.

Table~\ref{tab:web-operational-diagnostics} reports the retained Web outcomes by selected attack channel.

\begin{table}[H]
\centering
\scriptsize
\caption{Web outcomes by selected attack channel ($n=112$). Green and pink mark the highest and lowest RSR.}
\label{tab:web-operational-diagnostics}
\setlength{\tabcolsep}{2.2pt}
\renewcommand{\arraystretch}{1.04}
\begin{tabular*}{\columnwidth}{@{\extracolsep{\fill}}lcc@{}}
\toprule
\textbf{Channel} & \textbf{Success} & \textbf{RSR} \\
\midrule
 Browser            & 29/30 & \cellcolor{green!18}\textbf{96.7\%} \\
 Hybrid             & 23/30 & 76.7\% \\
 Protocol           & 26/34 & 76.5\% \\
 Browser + protocol & 13/18 & \cellcolor{red!12}72.2\% \\
\bottomrule
\end{tabular*}
\end{table}

\begin{takeaways}
\textbf{Answer to RQ4.}
\textit{Level~2 identifies \textbf{93/112} cases as state-dependent, of which TAILOR reproduces \textbf{79}. All \textbf{19} multi-actor cases with retained gate results enter scenario planning, showing that Level~2 adapts coordination to state complexity.}
\end{takeaways}
\FloatBarrier

\vspace{-0.35em}
\subsection{Effect of Level 2 State Awareness}
\label{sec:eval-state-effect}
\vspace{-0.35em}

Section~\ref{sec:eval-stateful} shows that Level~2 state awareness can convert the state requirements of Web CVEs into constraints for exploitation and verification. RQ5 further examines its effect on the end-to-end reproduction of complex Web CVEs. We conduct a No-L2 ablation on 10 stateful Web CVEs that TAILOR successfully reproduces. For each case, No-L2 starts from the original CVE input and reruns the complete workflow; Appendix~\ref{app:ablation-cases} reports the case-level results.

\noindent\textbf{No-L2 configuration.} No-L2 retains the Web workflow and its execution tools and only removes the coordination provided by Level~2 across attack preparation, exploitation, and verification. The system no longer provides preconstructed state or scenario constraints; the attack channel uses a generic strategy, and verification does not use the corresponding state-differential conditions.

\begin{figure}[t]
\centering
\duetbeginsubfigures
\makebox[\columnwidth][c]{%
  \hspace*{-0.064\columnwidth}%
  \includegraphics[width=0.68\columnwidth]{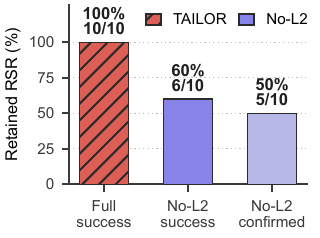}%
  \hspace*{0.064\columnwidth}%
}
\duetsubcaption{Outcome retention.}
\label{fig:no-l2-outcome}
\vspace{0.1em}
\includegraphics[width=0.96\columnwidth]{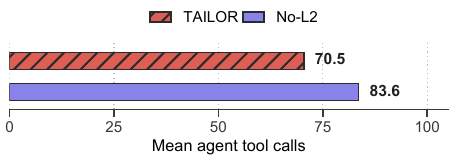}
\duetsubcaption{Agent tool use.}
\label{fig:no-l2-operational}
\duetfinishsubfigures
\caption{No-L2 evaluates the effect of Level~2 on verified outcomes and execution behavior.}
\label{fig:no-l2-ablation}
\end{figure}

Figure~\ref{fig:no-l2-ablation} shows that the full workflow reaches V3 confirmation for all 10 cases. Under No-L2, six cases are reported as successful, but only five receive V3 confirmation. The remaining CSRF case preserves only a direct-request result. It lacks the browser context needed to demonstrate the vulnerability effect and therefore does not qualify as a verified (V3) reproduction. Of the other cases, one fails during exploitation and one produces insufficient evidence. The remaining two terminate during environment construction and after reaching the cost limit, respectively. After Level~2 is removed, mean Agent tool calls increase from 70.5 to 83.6, while mean Verifier confidence decreases from 0.94 to 0.83. Without explicit state preparation and scenario constraints, the Agent therefore performs more exploration, and its evidence is less likely to satisfy verification.

\begin{takeaways}
\textbf{Answer to RQ5.}
\textit{On the selected stateful Web CVEs, No-L2 retains \textbf{6/10} self-reported successes, but only \textbf{5/10} receive V3 confirmation. Full TAILOR confirms 10/10. Removing Level~2 also increases tool use, showing that this mechanism is important for retaining verifiable end-to-end reproductions.}
\end{takeaways}

\subsection{Result Validation and Cost}
\label{sec:eval-cost}

The preceding experiments evaluate TAILOR's reproduction effectiveness and the roles of its two-level design. This section further analyzes the reliability of the final decisions and the operating cost of the full system.

\noindent\textbf{Result validation.} Among \textbf{85} cases that retain both an Agent self-report and a Verifier verdict, the Verifier rejects \textbf{5} Agent-claimed successes. Independent replay reobserves \textbf{36/51} claimed vulnerability effects. Three additional replay attempts are withheld because their PoC scripts contain unsafe shell commands. These results show that most replayed successes can be independently confirmed. Appendix~\ref{app:ghost-case} provides a complete example of an evidence-based decision.

\noindent\textbf{Cost.} The mean cost per CVE is \textbf{\$2.79}, the median is \textbf{\$1.34}, and the 90th percentile is \textbf{\$7.70}. Only \textbf{5/155} runs (3.2\%) exceed their route-specific cost cap, and only after evidence has already been confirmed.

\section{Limitations and Future Work}
\label{sec:limitations}

\noindent\textbf{Coverage across CVE classes.}
TAILOR's current Web/traditional dichotomy is not a complete taxonomy of CVEs. Hardware, IoT, and mobile vulnerabilities may depend on device-specific interfaces, firmware, emulators, communication protocols, or physical peripherals, none of which is explicitly modeled by either path. A direct extension is therefore to generalize the type-aware controller from binary routing to multi-class routing. Each class then defines its own environment requirements, trigger interface, and execution stages.

\noindent\textbf{Windows-specific targets.}
The execution backend relies on Linux-based Docker environments. It therefore cannot cover Windows-only CVEs whose triggers depend on Windows APIs, services, kernel behavior, or drivers. Supporting these cases requires a Windows-native execution backend, such as isolated virtual machines or Windows containers, together with artifact-preservation mechanisms for host-specific dependencies. This extension is not merely adding a routing label. A Windows path must reproduce the runtime semantics required by the target.

\noindent\textbf{Agent tool autonomy.}
TAILOR's agents work only through a fixed set of tools. When a CVE requires an interaction outside this set, an agent may recognize the missing step but cannot execute it. Future work will expand browser, system, and device interaction support, and allow agents to select, compose, and parameterize tools from runtime feedback. Greater autonomy must remain bounded by isolated targets, action logging, and explicit safety policies.

These extensions will generalize TAILOR's two-path design into a broader type-aware reproduction framework while preserving its central principle: each CVE class should be reproduced through execution machinery that matches its runtime form and trigger interface.

\FloatBarrier

\section*{Ethics Considerations}

\noindent\textbf{Stakeholders and benefits.} The direct stakeholders are maintainers and users of the evaluated software, security researchers and defenders who may reuse the benchmark, artifact evaluators, and model-service providers that process public vulnerability context. The intended benefits are faster validation of public CVEs, more auditable security benchmarks, and evidence that helps defenders test mitigations. The evaluated CVEs were public before this study, and TAILOR was not used to discover or scan for undisclosed vulnerabilities.

\noindent\textbf{Risks and mitigations.} Automated reproduction can lower the barrier to weaponizing public vulnerability information and produce dual-use exploit artifacts~\cite{alalsadi2026dualuse}. We limit experiments to locally controlled Docker deployments and do not intentionally direct exploit traffic at third-party or production services. TAILOR separates exploitation from independent verification, rejects unsupported evidence, and constrains replay to the intended target. Replay refuses unsafe shell commands; three requests during validation were therefore not executed. Before release, we remove credentials and transient session material and review each PoC for operational risk. A per-case artifact may be withheld when publication would create excessive risk, with the reason recorded in the release materials. These measures reduce, but do not eliminate, the risk that released automation is repurposed offensively~\cite{menlo2012report}.

\noindent\textbf{Data and disclosure.} The study uses only public vulnerability records, advisories, patches, and source code. It does not recruit human participants or intentionally collect personal data. Inputs sent to model services are limited to public vulnerability information and synthetic or locally generated execution state. If future use of TAILOR reveals an undisclosed issue, we will follow coordinated disclosure before releasing the corresponding exploit material.

\bibliographystyle{IEEEtran}
\bibliography{\jobname}

\raggedbottom
\appendices

\section{Reproduction Evidence Walkthrough}
\label{app:ghost-case}

\noindent\textbf{CVE-2024-43409.} This case is an unauthenticated newsletter-unsubscribe vulnerability in Ghost CMS (CWE-287/639). The exploit looks deceptively simple, but verifying its evidence is considerably harder. The unsubscribe endpoint trusts the member UUID embedded in the link and requires nothing else: no session, no cookie, and no HMAC key. Anyone who obtains the victim's UUID can therefore silently change that member's subscription state, and TAILOR's PoC is a single unauthenticated GET request. The server answers this request with a 302 redirect to the portal's unsubscribed view, but the redirect neither reports nor modifies any member state. A verifier that reads only the exploit's own response would therefore count a ``nothing happened'' event as a success. TAILOR does not rely on this response. Instead, it reads the victim's record through the admin API once before the request and once after (Figure~\ref{fig:case-cve-2024-43409}). The two reads differ in exactly one place: the newsletter list changes from active to empty, and the record's \texttt{updated\_at} timestamp is refreshed accordingly. Because the request carries no session or signature material, no other actor and no prior state can explain this change; the unauthenticated trigger is the only explanation. The independently observed cross-actor state transition, rather than the 302, is thus the V3 evidence that closes the case, establishing the causal link from the unauthenticated trigger to the victim's state change.

\begin{figure}[H]
\centering
\begin{casebox}{Interaction-based PoC\,---\,single GET, no credentials}
\ttfamily\scriptsize\linespread{0.9}\selectfont
\noindent
GET /members/unsubscribe/\par
\hspace*{1.2em}?uuid=b1f79416-3b38-4ee3-a63b-e4a7c8d22fe4\par
\hspace*{1.2em}\&redirect=http://localhost:2368/\par
Host: localhost:2368\par
\smallskip
\hlR{(no Cookie\,$\cdot$\,no HMAC key\,$\cdot$\,UUID is the only credential)}\par
\smallskip
$\rightarrow$ HTTP/1.1 302 Found\par
Location: /\#/portal/unsubscribed?uuid=...\&\hlY{key=}\par
\hspace*{5.4em}{\rmfamily\scriptsize\hlY{empty}\,---\,no key was issued or checked}\par
\end{casebox}

\smallskip

\begin{casebox}{Replay-gate Cross-state Evidence\,---\,admin-API probe before/after}
\ttfamily\scriptsize\linespread{1.05}\selectfont
\noindent
\textbf{BEFORE}\quad GET /ghost/api/admin/members/671182b2\dots a8e10/ $\rightarrow$ 200\par
\hspace*{1em}"newsletters": [\{"id":"641f5ee5\dots a4e66d",\par
\hspace*{4.6em}"name":"The Weekly Ghost",\par
\hspace*{4.6em}\hlG{"status":"active"}\}]\par
\smallskip
\textbf{AFTER}\quad\, GET /ghost/api/admin/members/671182b2\dots a8e10/ $\rightarrow$ 200\par
\hspace*{1em}\hlG{"newsletters": []}\par
\hspace*{1em}"updated\_at": \hlY{"2024-10-17T23:16:07.000Z"}\par
\smallskip
$\Rightarrow$ \hlG{Cross-actor state diff confirmed}\,---\,victim record\par
\hspace*{1em}mutated by an unauthenticated request\par
\hspace*{1em}(independent differential probe).\par
\smallskip
\end{casebox}\par
\caption{Accepted Ghost CMS reproduction. V3 evidence is the before/after admin probe, not the unauthenticated exploit redirect.}
\label{fig:case-cve-2024-43409}
\end{figure}

\section{Dataset Construction and Selection Details}
\label{app:dataset-composition}

\subsection{Dataset Construction}

\noindent\textbf{Source A.} Source~A contains 89 CVEs from the CVE-GENIE benchmark~\cite{ullah2025cvegenie}: 58 Web and 31 traditional cases. This source provides a shared benchmark for comparison with prior work.

\noindent\textbf{Source B.} Source~B contains 111 CVEs independently collected from public CVE records and security advisories: 99 Web and 12 traditional cases. Collection emphasized open-source Web components, but also retained in-scope traditional cases. We deduplicated Source~B against Source~A by CVE ID.

\begin{center}
\footnotesize
Candidate CVEs $\rightarrow$ identifiable affected version $\rightarrow$ source/archive available
\par$\rightarrow$ current execution scope $\rightarrow$ remove Source-A overlap $\rightarrow$ final Source~B
\end{center}

\subsection{Inclusion and Exclusion Criteria}

Table~\ref{tab:dataset-criteria} summarizes the criteria used before running TAILOR. Selection did not depend on whether TAILOR could reproduce a candidate.

\begin{table}[H]
\centering
\scriptsize
\caption{Source-B selection criteria.}
\label{tab:dataset-criteria}
\setlength{\tabcolsep}{4pt}
\begin{tabular}{p{0.66\columnwidth}p{0.22\columnwidth}}
\toprule
\textbf{Criterion} & \textbf{Decision} \\
\midrule
Publicly disclosed CVE & Include \\
Affected version identifiable & Required \\
Source code or version archive available & Required \\
Public PoC available & Not required \\
CVE ID overlaps with Source~A & Exclude \\
Required reconstruction inputs unavailable & Exclude \\
Outside the current Web/traditional execution scope & Exclude \\
\bottomrule
\end{tabular}
\end{table}

\subsection{Dataset Composition}

Table~\ref{tab:dataset-source-composition} gives the source and type composition of the final 200-CVE corpus.

\begin{table}[H]
\centering
\scriptsize
\caption{Dataset composition by source and CVE type.}
\label{tab:dataset-source-composition}
\setlength{\tabcolsep}{5pt}
\begin{tabular}{lrrr}
\toprule
\textbf{Source} & \textbf{Web} & \textbf{Traditional} & \textbf{Total} \\
\midrule
Source A & 58 & 31 & 89 \\
Source B & 99 & 12 & 111 \\
\midrule
\textbf{Total} & \textbf{157} & \textbf{43} & \textbf{200} \\
\bottomrule
\end{tabular}
\end{table}

We canonicalized repository URLs case-insensitively, resolved verified transfers and renames, and preserved independently released plugin and module boundaries. This audit mapped 161 raw keys extracted from repository text to \textbf{160 normalized projects}; every project-identity candidate received an explicit counting decision. Table~\ref{tab:dataset-composition} reports the corpus's primary-language composition under both CVE-weighted and project-weighted counting.

\begin{table}[H]
\centering
\scriptsize
\caption{Audited project and primary-language composition. Each CVE contributes once to the CVE column; each normalized project contributes once to the project column.}
\label{tab:dataset-composition}
\setlength{\tabcolsep}{3.2pt}
\begin{tabular}{lrr@{\hspace{7pt}}lrr}
\toprule
\textbf{Language} & \textbf{CVEs} & \textbf{Projects} &
\textbf{Language} & \textbf{CVEs} & \textbf{Projects} \\
\midrule
PHP        & 40 & 24 & Rust       & 6 & 5 \\
Python     & 35 & 27 & C\#        & 3 & 3 \\
TypeScript & 30 & 25 & C++        & 3 & 3 \\
Go         & 28 & 23 & HTML       & 2 & 2 \\
JavaScript & 20 & 18 & Elixir     & 1 & 1 \\
Java       & 12 & 11 & Kotlin     & 1 & 1 \\
C          & 11 &  9 & Shell      & 1 & 1 \\
Ruby       &  6 &  6 & Vim Script & 1 & 1 \\
\midrule
\multicolumn{1}{l}{\textbf{Total}} & \textbf{200} & \textbf{160} &
\multicolumn{3}{r}{\textbf{16 primary languages}} \\
\bottomrule
\end{tabular}
\end{table}

The primary language is assigned from the target repository's language metadata, with structured deployment metadata and patch-file evidence used only as fallbacks. The distribution characterizes corpus composition rather than language-specific reproduction effectiveness.

\section{Implementation and Coordination Details}
\label{app:implementation}

\subsection{Routing-Rule Guardrails}
Table~\ref{tab:routing-rules} lists the representative routing-rule guardrails introduced in \S\ref{sec:routing}.

\begin{table}[ht]
\centering
\scriptsize
\caption{Representative routing-rule guardrails layered over the LLM classifier.}
\label{tab:routing-rules}
\setlength{\tabcolsep}{4pt}
\begin{tabular}{lll}
\toprule
\textbf{Indicator type} & \textbf{Example signals} & \textbf{Route} \\
\midrule
CWE class            & CWE-89, 79, 352, 918, 284/285/863       & Web \\
Web framework        & Flask, Django, FastAPI, Express,        & Web \\
                     & Spring Boot, WordPress, Drupal           & \\
Patch file path      & \texttt{controller/}, \texttt{routes/}, \texttt{api/}, & Web \\
                     & \texttt{views/}, \texttt{templates/}    & \\
\midrule
Package distribution & npm, PyPI, NuGet ``package''            & Trad. \\
Logic-flaw markers   & ``incorrect calculation,''               & Trad. \\
                     & ``predictable,'' ``presigned URL''       & \\
Embedded / IoT       & router, firmware, IoT, PLC, SCADA       & Trad. \\
\bottomrule
\end{tabular}
\end{table}

\subsection{Agent Roster and Model Allocation}
\label{sec:roster}
This appendix catalogs the agents referenced throughout the design (\S\ref{sec:design}) and the model assigned to each.
TAILOR organizes its stage-specific agents into five functional clusters---pre-analysis, environment construction, exploitation and reflection, verification, and diagnostic recovery---supported by several rule-based or single-LLM-call helpers (Table~\ref{tab:model-allocation}). The allocation mixes LLM-driven agents (open-ended outputs) with rule-based controllers (structured, replay-critical outputs) along the dual-path pipeline.

\begin{table}[ht]
\centering
\scriptsize
\caption{Component roster by cluster, pipeline, and execution style.}
\label{tab:model-allocation}
\resizebox{\columnwidth}{!}{
\begin{tabular}{llcl}
\toprule
\textbf{Component} & \textbf{Model} & \textbf{Pipeline} & \textbf{Execution Style} \\
\midrule
Vulnerability Classifier & Claude Haiku 4.5    & Shared & Structured classification \\
Pipeline Router          & Rule / controller   & Shared & Deterministic routing \\
\midrule
Pre-analysis (5 LLM agents)
                         & Claude Sonnet 4.6   & Shared & Tool-augmented generation \\
\midrule
Web env construction (6 LLM agents)
                         & Claude Sonnet 4.6   & Web    & ReAct + lightweight planning \\
Trad env construction (RepoBuilder + critic)
                         & Claude Sonnet 4.6   & Trad.  & ReAct + lightweight planning \\
Deployment helpers (Planner + Executor)
                         & Rule / one-shot LLM & Shared & Programmatic safety net \\
\midrule
WebDriverAgent + critic  & Claude Sonnet 4.6   & Web    & ReAct (browser/protocol/hybrid) \\
Exploiter + critic       & Claude Sonnet 4.6   & Trad.  & ReAct-style execution \\
Reflectors (3 LLM agents)
                         & Claude Sonnet 4.6   & Shared & Cross-iteration analysis \\
\midrule
WebVerifier              & Rule / adapter      & Web    & Evidence-driven \\
CTFVerifier + SanityGuy  & Claude Sonnet 4.6   & Trad.  & Verifier + sanity audit \\
\midrule
Diagnostic \& recovery (3 components)
                         & Sonnet 4.6 + rules  & Shared & Failure remediation \& budget cut-off \\
\midrule
Artifact Manager         & Rule / controller   & Shared & Structured persistence \\
\bottomrule
\end{tabular}
}
\end{table}

\noindent\textbf{Cluster contents.}
\emph{Pre-analysis} (5 agents producing Fig.~\ref{fig:preanalysis-schema}): \emph{CVEInfoGenerator}, \emph{KnowledgeBuilder}, \emph{PreReqBuilder}, \emph{DeploymentStrategyAnalyzer}, and \emph{ConfigInferencer}.
\emph{Environment construction}: Web---\emph{WebEnvBuilder}, \emph{ProjectSetupAgent}, \emph{ServiceStartAgent}, \emph{HealthCheckAgent}, \emph{StateInitializerAgent}, and the post-completion review of hybrid evidence by \emph{WebEnvCritic} (\S\ref{sec:env}); traditional---\emph{RepoBuilder} (dispatching to \emph{ServiceRepoBuilder}\,/\,\emph{LibraryRepoBuilder}), reviewed by \emph{RepoCritic}; shared---\emph{DeployPlanner} and \emph{DeployExecutor}.
\emph{Exploitation \& reflection}: \emph{WebDriverAgent} and \emph{Exploiter} handle Web and traditional execution, respectively; \emph{ExploitCritic} operates only on the traditional side (the Web side delegates this role to verification, \S\ref{sec:exploit-exec}); \emph{ExploitReflector} supplies cross-iteration experience records---verified facts, failed attempts, blockers, and next strategies---to environment construction and Web exploitation (\S\ref{sec:env}, \S\ref{sec:exploit-exec}).
\emph{Verification}: \emph{WebVerifier} (rule-based adapter), \emph{CTFVerifier} with \emph{HardenedVerifier} program logic, and \emph{SanityGuy}, which audits the verifier itself.

\noindent The roster follows two allocation rules: LLM-driven where outputs are open-ended (build, exploit, reflect), and rule-driven where outputs are structured and replay-critical (verification, deployment helpers, recovery)---trading determinism for case-coverage breadth, and coverage for verification-path auditability.

\makeatletter
\setlength{\@dblfptop}{0pt}
\setlength{\@dblfpsep}{8pt}
\setlength{\@dblfpbot}{0pt plus 1fil}
\makeatother
\begin{table*}[!t]
\centering
\scriptsize
\caption{Default Web-exploitation settings indexed by CWE family.}
\label{tab:cwe-coordination}
\setlength{\tabcolsep}{4pt}
\resizebox{\textwidth}{!}{%
\begin{tabular}{lllll}
\toprule
\textbf{CWE family} & \textbf{Default channel} & \textbf{Scenario gate} & \textbf{State init.\ contract} & \textbf{Verification oracle} \\
\midrule
CWE-79 (XSS)
  & browser / hybrid
  & usually false; true if stored or multi-user
  & browser DOM / session
  & DOM canary, alert, screenshot, network event \\
\addlinespace
CWE-352 (CSRF)
  & browser / hybrid
  & often true (victim state change)
  & victim session state
  & before/after state diff \\
\addlinespace
CWE-918 (SSRF)
  & protocol
  & usually false
  & request / OOB state
  & internal response, DNS/HTTP callback \\
\addlinespace
CWE-89 (SQLi)
  & protocol
  & false
  & request / response
  & DB error, timing, extracted rows \\
\addlinespace
CWE-78/94 (cmd inj.\ / RCE)
  & protocol
  & false
  & request / canary
  & command output, file canary, OOB \\
\addlinespace
CWE-22/98 (traversal / LFI)
  & protocol
  & usually false
  & request / session if auth
  & file-content marker \\
\addlinespace
CWE-284/287/306/639/862/863
  & hybrid / protocol
  & true
  & actor--resource--policy / multi-actor
  & unauthorized access + negative control \\
(broken access control) & & & & \\
\addlinespace
CWE-434 (file upload)
  & hybrid
  & conditional
  & upload artifact state
  & served file, execution marker \\
\bottomrule
\end{tabular}
}
\end{table*}

\subsection{CWE-indexed Web-exploitation Defaults}
\label{app:cwe-coordination}

Table~\ref{tab:cwe-coordination} lists the per-CWE defaults referenced in \S\ref{sec:webexp}. Each row gives the channel the exploit typically uses (protocol, browser, or both), whether scenario planning runs first, what state preparation supplies to downstream stages, and what the verifier looks for as success evidence. These are starting points, not hard-coded rules: each module overrides its default when runtime evidence conflicts with it.

\section{Ablation Case Details}
\label{app:ablation-cases}

\subsection{No-L1 Ablation Cases}

The No-L1 set contains ten Web CVEs selected from cases that full TAILOR had reproduced with V3 evidence. We kept the CVE input fixed and forced each case through the legacy workflow. This success-conditioned stress set measures retention under path mismatch; it is not a corpus-wide success estimate. The two legacy acceptances certify success with weaker local or static oracles than the V3 evidence the Web path produced, so 2/10 is an upper bound on retention (Table~\ref{tab:ablation-case-details}(a)).

\subsection{No-L2 Ablation Cases}

The No-L2 set contains ten explicitly state-dependent Web CVEs that full TAILOR reproduced with V3 evidence. No-L2 reruns each original CVE input without Level-2 state preparation, scenario coordination, channel prior, or state-differential verification constraints. Table~\ref{tab:ablation-case-details}(b) separates the Agent result from final verification; six runs report success, but only five reach V3. ``Rejected'' means the run reached the verifier but did not obtain V3 confirmation; ``Not reached'' means it terminated earlier.

\begin{table*}[!t]
\centering
\scriptsize
\caption{Per-case outcomes for the two success-conditioned ablation sets. Full denotes the full TAILOR result.}
\label{tab:ablation-case-details}
\textbf{(a) No-L1: forced legacy workflow}\par\smallskip
\setlength{\tabcolsep}{3.4pt}
\begin{tabular}{p{1.05in}p{1.20in}p{0.55in}p{0.48in}p{0.65in}p{1.35in}}
\toprule
\textbf{CVE} & \textbf{Project} & \textbf{CWE} & \textbf{Full} & \textbf{No-L1} & \textbf{Terminal stage} \\
\midrule
CVE-2024-10366 & LibreChat & 284 & V3 & Fail & Env. construction \\
CVE-2024-10719 & phpIPAM & 79 & V3 & Fail & Env. construction \\
CVE-2024-55661 & Laravel Pulse & 94 & V3 & Accepted* & Legacy verifier \\
CVE-2024-56800 & Firecrawl & 918 & V3 & Fail & Env. construction \\
CVE-2025-22617 & WeGIA & 79 & V3 & Fail & Env. construction \\
CVE-2025-22619 & WeGIA & 79 & V3 & Fail & Env. construction \\
CVE-2025-25300 & smartbanner.js & 601 & V3 & Accepted* & Legacy verifier \\
CVE-2025-27108 & dom-expressions & 79 & V3 & Fail & Env. construction \\
CVE-2025-27420 & WeGIA & 79 & V3 & Fail & Prerequisites \\
CVE-2025-27499 & WeGIA & 79 & V3 & Fail & Prerequisites \\
\bottomrule
\end{tabular}
\par\vspace{6pt}
\textbf{(b) No-L2: Level-2 coordination removed}\par\smallskip
\setlength{\tabcolsep}{3.1pt}
\begin{tabular}{p{1.05in}p{1.05in}p{0.48in}p{1.48in}p{0.42in}p{0.62in}p{1.05in}}
\toprule
\textbf{CVE} & \textbf{Project} & \textbf{CWE} & \textbf{State dependency} & \textbf{Full} & \textbf{Agent} & \textbf{Verification} \\
\midrule
CVE-2025-64523 & File Browser & 285 & Multi-actor ownership & V3 & Failure & Not reached \\
CVE-2025-68481 & FastAPI Users & 285 & Cross-session OAuth state & V3 & Failure & Rejected \\
CVE-2024-47053 & Mautic & 285 & Auth. resource ownership & V3 & Success & V3 confirmed \\
CVE-2025-48371 & OpenFGA & 285 & Relation-policy state & V3 & Success & V3 confirmed \\
CVE-2025-62610 & Hono & 285 & Signed-token claims & V3 & Failure & Not reached \\
CVE-2025-30373 & Graylog & 285 & Message-processing state & V3 & Failure & Not reached \\
CVE-2025-53092 & Strapi & 200 & Session/origin context & V3 & Success & V3 confirmed \\
CVE-2026-44364 & MISP modules & 352 & Browser session/CSRF & V3 & Success & Rejected \\
CVE-2026-30224 & OliveTin & 384 & Session lifecycle & V3 & Success & V3 confirmed \\
CVE-2026-27167 & Gradio & 522 & Server session/cookie & V3 & Success & V3 confirmed \\
\bottomrule
\end{tabular}
\end{table*}

\section{Trace Interface View}
\label{app:trace-ui}

\begin{figure}[H]
\centering
\includegraphics[width=0.80\columnwidth]{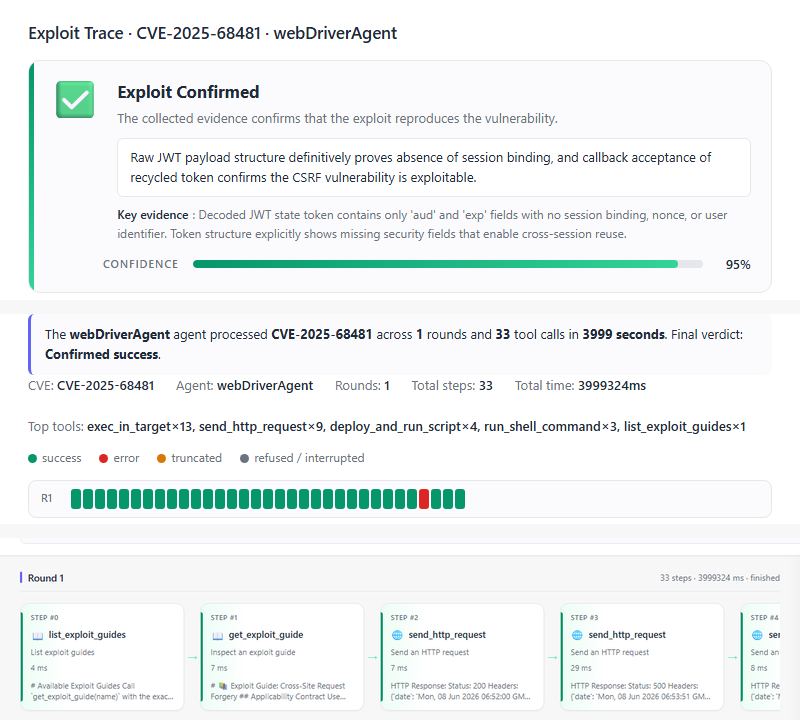}
\caption{Static trace-interface view for CVE-2025-68481. The retained trace records a 95\% confirmed verdict after 33 tool calls; the red node marks a recovered intermediate error.}
\label{fig:trace-ui}
\end{figure}

\noindent\textbf{CVE-2025-68481.} Figure~\ref{fig:trace-ui} shows a compact static rendering of the trace interface retained with the case artifact. The view combines the evidence-backed verdict, execution summary, per-step status density, and the first trace nodes. The original artifact remains an interactive HTML document; this figure preserves its audit-oriented presentation in the paper format.

\end{document}